\documentstyle[11pt,aasms4]{article}
\begin{document}

\title{Infrared Variation of Blazars}

\author{J.H. Fan}

\affil{
Center for Astrophysics, Guangzhou Normal University, Guangzhou 510400, 
 China, e-mail: jhfan@guangztc.edu.cn}
\date{Received <date>;accepted<date>}

\begin{abstract}

 In this paper, the historical infrared (JHK) data compiled from 
 the published literature are presented here for  30 blazars. Maximum
 near-IR variations are found  and compared with the  optical ones.
 Relations 
 between color index and magnitude  and between color-color indices are 
 discussed individually. For the color-magnitude relation, some objects 
 (0215+015, 0422+004, and 1641+395)  show that the color index 
 increases with magnitude indicating that the 
 spectrum flattens when the source brightens while 1253-055
 shows complex behaviour: the spectrum steepens when the source 
 dims in J band and the spectrum flattens when the source dims
 further suggesting that the emission mechanism consists of, 
 at least, two parts in this case. From the color indexes,
 we have that the spectral indexes are in the range of 
 $\alpha_{IR} = 0.77 \sim 2.37$ ($f_{\nu} \propto \nu^{-\alpha}$).
 
 \end{abstract}

 \section{Introduction}

 While the nature of active galactic nuclei (AGNs) is still an open problem, 
 the study of AGNs variability  can yield valuable information about 
 their nature.  Photometric observations of AGNs are important to 
 construct light curves and to study  variability behavior over
 different time scales. In AGNs, some long-term optical variations have been 
 observed and in some cases claimed to be periodic (see Fan et al. 1998a). 

 Blazars are AGNs characterized by compact radio core, high and variable 
 radio and optical polarization, superluminal radio components. The 
 continuum emissions are rapidly variable at all frequencies with amplitude 
 and rapidity increasing with frequency ( see Kollgaard 1994 ). Blazars 
 include BL Lac objects, optically violently variable quasars (OVVS), 
 highly polarized quasars (HPQs), flat spectrum radio quasars (FSRQ) and 
 core dominated quasars (CDQ). All those objects are basically the same 
 thing (Fugmann 1989; Impey et al. 1991; Valtaoja et al. 1992; 
 Will et al. 1992). 

 Before $\gamma$-ray observations were available, Impey \& Neugebauer (1988) 
 found that the infrared emission  (1-100 $\mu m$) dominates the bolometric 
 luminosities of blazars. The infrared emissions are also an  important 
 component for the luminosity even when the $\gamma$-ray emissions are 
 included ( von Montigny 1995). Study of the infrared will provide much 
 information of the emission mechanism. The long-term infrared variations 
 have been  presented for some selected blazars in a paper by 
 Litchfield et al. (1994). 

 Infrared observations have been done for blazars for more than 20 years. 
 Neugebauer et al. (1979) presented the infrared variations for a few 
 blazars with the observation time being back to the 1960's.  The infrared 
 observations done over a period of 8 years have also been presented for 
 some selected blazars in the paper by Litchfield et al. (1994).  But there 
 are no available long-term infrared variations in the literature for all 
 these objects. Recently, we have obtained the long-term infrared variation 
 for 40 radio selected BL Lac objects (Fan \& Lin 1999a). In this paper, 
 we mainly present the long-term infrared (J,H, and K bands ) light curves 
 for some blazars and discuss the variation properties. The compiled data 
 available from the author (e-mail: jhfan@guangztc.edu.cn).
 This paper has been arranged similar to our previous paper (Fan \& Lin
1999a).
  In section 2, we review the 
 literature data and  light curves; in section 3, we discuss the 
 variations and give some remarks for each object,  and a short  summary.

\section{Near-infrared Light Curves}

\subsection{Data from Literature }

 Infrared observations are available since the end of the 1960s. We 
 compiled the data  from 47 publications. They are listed in table 1 \& 2,
  which gives the observers in Col. 1.  and the telescope(s) used in Col.2. 
 For the J, H, and K magnitudes, they are listed in an table 4 in
 electronic form. In table 4, Col. 1 gives the name; Col. 2. the JD time;
 Col. 3, the J magnitude; Col. 4 the uncertainty for J; Col. 5, the 
 H magnitude; Col. 6, the uncertainty for H; Col. 7, K magnitude;
 Col. 8, the uncertainty for K.

\subsection{Data Analysis}

 The flux densities from the literature had been converted back to magnitudes
 using the original formulae.  In the literature, different telescopes 
 are used, different telescopes use photometers with slightly
 different filter profiles, resulting in slightly  different calibration 
 standards and zero-points, but the  uncertainty aroused by the different 
 systems is no more than a few percent.  

 The magnitude is dereddened using  
 $A(\lambda)=A_{V}(0.11\lambda^{-1}+0.65\lambda^{-3}-0.35\lambda^{-4})$
 for $\lambda > 1 \mu m$ and  $A_V$ = 0.165(1.192- $\tan{b})\csc{b}$ for 
 $ |b| \leq 50^{\circ}$ and $A_V$ = 0.0 for  $ |b| > 50^{\circ}$ 
 (Cruz-Gonzalez \& Huchra 1984, hereafter CH (84); Sandage 1972; 
 Fan et al. 1998b).  It is clear that some objects have many observations 
 while others have only a few data points, in Figures 1-15, we  show only 
 those with  more observations.

 For data shown in Figs. d - i of each object we have
 performed linear fit with uncertainties  in both coordinates 
 considered (see Press et al. 1992 for detail), 
 \begin{equation}
 y(x) = a + bx
 \end{equation}
 In principle, $a$ and $b$ can be determined  by 
 minimizing the $\chi^{2}$ merit function (i.e. equation 15.3.2 in 
 Press et al. book)
 \begin{equation}
\chi^{2}(a,b)= {\sum\limits_{i=1}^{n} {(y_{i}-a-bx_{i})^{2}w_{i} }} 
 \end{equation}
 where ${\frac{1}{w_{i}}}=\sigma^{2}_{y_{i}}+b^{2}\sigma^{2}_{x_{i}}$,  $\sigma_{y_{i}}$ and $\sigma_{x_{i}}$ are the $x$ and $y$ standard 
 deviations for the $i$th point
 Unfortunately, the occurrence of $b$ in the denominator of the above 
 $\chi^{2}$ merit function, resulting in equation
 ${\frac{\partial \chi^{2}}{\partial b}}=0$ being
 nonlinear makes the  fit very complex, although we can get a 
 formula for $a$ from
  ${\frac{\partial \chi^{2}}{\partial a}}=0$:
 \begin{equation} 
  a={\frac{\sum\limits_{i}w_{i}(y_{i}-bx_{i})}{\sum\limits_{i}w_{i}}}
 \end{equation}
  Minimizing the $\chi^{2}$ merit function, equation (2), with respect to 
 $b$ and using the equation (3) for $a$ at each stage to ensure that the 
 minimum with respect to $b$ is also minimized with respect to $a$, we 
 can obtain $a$ and $b$. As Press et al. stated, the linear correlation 
 coefficient $r$ then can be  obtained by  weighting the relation (14.5.1)
  terms of Press et al. (1992).  
 Finding the  uncertainties, $\sigma_{a}$ and $\sigma_{b}$, in $a$ and $b$ is
 more complicated (see Press et al. 1992). Here, we have  not performed this.
 For the data whose uncertainties were not given in the original literature 
 are not included in our linear fit or in Fig. d-i, but they 
 are included in the light curves (Fig. a - c).

 As discussed in the paper of Massaro \& Trevese (1996) (also see Fan \&
 Lin 1999a), there is a 
 statistical bias in the  spectral index-flux density correlation. Following 
 their suggestion, we considered the relation between magnitude in one 
 band and the color index obtained from two other  bands to avoid this
 bias. The relations are shown in Fig. d-f.

\section{Discussion}

\subsection{Variations}

 Blazars are variable at all wavelengths. For blazars with enough
 infrared data, we presented their variability properties in Table 3, which 
 gives the name in Col. 1, redshift in Col. 2, reddening correction,
$A_{V}$,
 in Col. 3, largest optical polarization ($P_{opt}$) in Col. 4, maximum
 amplitude optical variation ($m_{opt}$) in Col. 5, maximum amplitude 
 variations in J, H, and K in Col. 6, 7, and 8, the averaged color index 
 (J-H) and (H-K) in Col. 9, and 10, the uncertainty in Col. 9  and 10 
 is the one sigma  deviation. The infrared variations are very large for 
 some objects. 
 For variation, it is reasonable for larger variation 
 to correspond to shorter wavelength, but some objects show different 
 behaviour in the present paper. It is interesting to note that the object 
 having steepening spectrum when the source brightens shows that the 
 variation at the longer wavelength  is larger than that at the shorter 
 one.  Most objects showing large optical variation also show
 large infrared variations and high polarizations, large variation and 
 high polarization are associated. For the color-indices, 
 we have that (J-H) covers a range of 0.66 to 1.44 suggesting a range
 of spectral index $\alpha$ = 0.77 to 2.37 since
 (J-H) = 0.300$\alpha$ + 0.43 (Sitko, private communication).

 It is possible that more than one mechanism is responsible for
 the emission in the infrared region, i.e., 
 it is reasonable that a nonvariable or
 slowly varying near-IR component, such as the stars in the parent galaxy,
 is present in the spectrum of AGNs.  In this sense, when the source is
 bright, the spectrum is observed to steepen when the source dims, as
 expected from a synchrotron component which experiencing radiative energy
 losses, but when the source dims further, because of the presence of
 the underlying near-IR emission, the spectrum will flattens with the
 source getting faint (Fan 1999a,b).  So, the investigation of the
relation
 between spectrum index and flux can throw some lights on the components
 of near-IR emission mechanism, particularly when the source is faint.
 Our analysis in the present paper indicates that there are perhaps
 at least two components in the near-IR region of 3C 279, but it 
 should be confirmed with more observations.

 Following the variation properties are discussed for individual object.

\subsection{Remarks}

\subsubsection{0109+224, $z=?$}
 This polarized ($P_{opt.}=17.26\pm1.63\%, P_{IR}=13.87\pm0.74\%$, 
 Mead et al. 1990) object shows a variation of $\Delta B=3.07$ (Bozyan et al. 
 1990). The infrared variations are less than  in  the optical one, 
 no correlation is found for color index and magnitude. There is an 
 indication of a correlation between (J-K) and (J-H), but not for other 
 color indices (see Fig. 1).

\subsubsection{0215+015, $z_{abs}=1.686$}
 
 It is one of the most luminous ($B=14.5 \sim \geq 19.5$ Pettini et al. 1983) 
 known BL Lac objects  and  polarized at $P_{IR}=17.4\pm1.1\%$ in H band 
 (Holmes et al. 1984) and $P_{opt}=20\%$ (Angel \& Stockman 1980). There 
 are only a few infrared data in the literature, which show a positive
  correlation between color index and magnitude suggesting the spectrum 
 flattens  when the source brightens: 
 (H-K) = 0.08J - 0.26 with $r$ = 0.97 and $p = 1.2\%$;
 (J-K) = 0.16H - 0.36 with $r$ = 0.89 and $p = 4.5\%$. Those color-magnitude
 correlations were obtained based on 4 points, so they should be confirmed with
  more observations. 
 For color indices, there is also an 
 indication of correlations between  (J-K) and (J-H) and (H-K) as well 
 with $p < 5\%$ (see Fig. 2). 

\subsubsection{0301-243, $z=?$}
 
 This moderately variable ($\Delta m=0.89$, Pica et al. 1988) object has 
 been observed in the infrared for 3 nights showing a variation of 0.20 mag 
 and color indices of (J-H) = 0.81$\pm$0.05, (J-K) = 1.52$\pm$0.03 and
 (H-K) = 0.71$\pm$0.05.  It is polarized at  $P_{opt.}=10.97\%$ 
 (Impey \& Tapia 1988).

\subsubsection{0323+022, $z=0.147$}

 It is one of the X-ray selected BL Lac objects (XBLs) showing extremely 
 rapid variation. An X-ray variation over a time scale of 30 seconds 
 was observed by Feigelson et al. (1986), who also noticed a brightness 
 decrease of 1.3 mag within one day in the optical region.  Polarizations 
 of $P_{opt.}\sim 2-9\%$ (Feigelson et al. 1986) and  $P_{IR}=4.65\pm1\%$
 ( Mead et al. 1990) are reported. It can be obtained that the optical 
 spectral index is strongly associated with the brightness 
 ( $V=-(1.52\pm0.04)(B-V)+17.07\pm0.02$, with a correlation coefficient 
 of $r=-0.967$) when the optical observations (Feigelson et al. 1996) are 
 taken into account. There are only a few infrared data for this object 
 showing variations of $\Delta J=0.30$, $\Delta H=0.52$ and $\Delta K=0.96$.

\subsubsection{0336-019, $CTA 26$, $z=0.852$}
 There are only 4 nights of infrared data for this polarized 
 ($P_{opt}=19.4\%$, Impey \& Tapia 1990)  and variable ($\Delta m=1.4$, 
 Pica et al. 1988) object showing J = 15.45, H = 15.00-15.96, and 
 K = 14.57-15.99  and color indices of
 (J-H) = 0.45$\pm$0.32, (J-K) = 0.88$\pm$0.29, and (H-K) = 0.43$\pm$0.25.

\subsubsection{0406+121, $z=1.02$}
 There are only 6 nights of infrared data for this faint (V = 20.4, 
 Rieke et al. 1979) object showing variations  of 
 $\Delta H = 0.90$, and $\Delta K = 1.78$ and color indices of 
 (J-H) = 1.13$\pm$0.07, (J-K) = 2.11$\pm$0.07 and (H-K) = 0.98$\pm$0.07.

\subsubsection{0420-014, $OA 129$, $z=0.915$}
 
 The infrared data show variations of $\Delta J = 2.46$, $\Delta H = 2.88$ 
 and $\Delta K = 2.61$ and color indices of (J-H) = 0.80$\pm$0.16, 
 (J-K) = 1.65$\pm$0.10, and (H-K) = 0.86$\pm$0.16. A polarization of 
 $P_{opt} = 20.19\pm1.26\%$ and an optical variation of $\Delta m$ = 2.8  
 are reported in the paper of Angel \& Stockman (1980). In addition, 
 the optical polarization is wavelength-dependent with polarization being 
 higher at longer wavelength (Smith et al. 1988). The limited data 
 indicate that K is  anti-correlated with (J-H), but this kind of 
 correlation is far from certainty.

\subsubsection{0422+004, $z=?$}
 
  An optical variation of $\Delta m=2.2$ (Branly et al. 1996) and high 
 and variable polarizations are known for this object. The infrared 
 polarization increases from $13.6\%$ to $19.8\%$ when the source
 dimmed by 0.3 mag in K (Holmes et al. 1984) and in other case it decreases 
 from $20.27\%$ to $12.25\%$ in two days when the  source does not show 
 significant variation, during which period the optical polarization also 
 decreases from  $22.14\%$ to $13.69\%$ (Mead et al; 1990). The infrared 
 data show that the infrared variations are  similar to the optical ones.
 A correlation between (H-K)  and J: (H-K) = -0.51 + 0.1J with  
 $r$ = 0.50 and $p = 4.4 \times 10^{-3}$. There are close correlations 
 between (J-K) and (J-H) and (H-K):
 (J-K) = 0.17 + 1.80(J-H) with $r$ = 0.81 and $p = 5.4 \times 10^{-8}$;
 (J-K) = 0.19 + 1.76(H-K) with $r$ = 0.789 and $p = 2.8 \times 10^{-7}$
 (see Fig. 3).

\subsubsection{0521-365, $z=0.055$}
 
 This steep-spectrum radio source was identified with an N galaxy by 
 Bolton et al. (1965b), a redshift of $z_{em}=z_{abs}=0.055$ was shown by 
 Danziger et al. (1978). A Polarization of $P_{opt} \sim 11\%$ 
 (Bailey et al. 1983) and a variation of $\Delta m =1.4$ (Angel \& Stockman, 
 1980) are reported. The infrared data show correlations between (J-K) and 
 (J-H) and (H-K) (see Fig. 4).
 (J-K) = 0.36 + 1.40(J-H) with $r$ = 0.89 and $p = 2.4 \times 10^{-4}$;
 (J-K) = -0.32 + 2.62(H-K) with $r$ = 0.756 and $p = 5.0 \times 10^{-3}$.
 The  infrared variation in K is similar to that in the optical band.

\subsubsection{0548-322, $z=0.069$}

 The radio source was found and identified with a 15 mag galaxy, photometry 
 yield V=15.5$\pm$0.1, (B-V) = 0.57$\pm$0.02, and  (U-B) = -0.30$\pm$0.03 
 (see Disney 1974).  The infrared data indicate that the variation in K is 
 larger than those in J and H, the color indices show an anti-correlation 
 between (J-H) and (H-K):
 (H-K) = 1.12  - 0.94(J-H) with $r$ = -0.76 and $p = 2.67\%$ (see Fig. 5)

\subsubsection{0716+322, $z=?$}
 
 The 4 nights of infrared  data show variations of $\Delta J=2.06$, 
 $\Delta H=2.05$ and $\Delta K=2.51$  and color indices of 
 (J-H) = 0.73$\pm$0.04, (J-K) = 1.50$\pm$0.02 and (H-K) = 0.84$\pm$0.12. 
 $P_{opt}=7.43\pm0.56\%$ and a wavelength-dependence of $P$, 
  $dP/d\lambda < 0$ are observed by Smith et al.(1988).

\subsubsection{0736+017, $OI+061$, $z=0.191$}

 This nearby quasar has  shown a variation of $\Delta m=1.35$ 
 (Pica et al. 1988). The optical polarization ($P_{opt}=0\%-6\%$) is less 
 than the infrared one ($P_{IR}=7.3\pm4.3\%$, Holmes et al. 1984). The 
 infrared light curves show that the source has been brightening with time, 
 but some brightness fluctuations also show up. (J-K) is found to be 
 correlated with  (J-H) and (H-K):
 (J-K) = 0.36 + 1.64(J-H) with $r$ = 0.83 and $p = 1.2 \times 10^{-3}$
 (J-K) = -0.11 + 2.04(H-K) with $r$ = 0.73  and $p = 7.0 \times 10^{-3}$.
  Complex associations can be seen 
 for color index and magnitude in Fig. 6,e,f (see Fig. 6).

\subsubsection{0823-223, $z > 0.91$}
 
 This polarized ( $P_{opt}=4\%-10.75\%$, Impey \& Tapia 1988) object has  
 shown an optical variation of $\Delta V=1.5$ and (V-K) = 3.7. The optical 
 spectrum steepens when the source brightens ($\alpha_{opt}=1.7\pm0.02$
  when V=15.7, and  $\alpha_{opt}=2.1\pm0.1 $ when V=16.6, Falomo 1990). 
 It has been observed in the infrared on some occasions, the data  show 
 an indication of an anti-correlation between color index and magnitude, 
 which is different from the optical behaviour:
 (J-H) =  3.19 - 0.20K with $r$ = -0.87  and $p = 1.2\%$.
 For the color indices, 
 (J-K) is correlated with (J-H) and (H-K) (see Fig. 7).

\subsubsection{0912+297, $OK222$, $z=?$}

 There are only 6 nights of infrared data for this polarized 
 ($P_{IR}=13.5\sim2.8\%$, Holmes et al. 1984, $P_{opt}=13 \%$, 
 Angel \& Stockman 1980) object showing variations 
 of  $\Delta H=2.27$, and $\Delta K=2.41$ and color indices of 
 (J-H) = 0.71$\pm$0.08, (J-K) = 1.53$\pm$0.11 and (H-K) = 0.85$\pm$0.20.  
 The optical variation ( $\Delta m=2.25$, Bozyan et al. 1990) is similar 
 to in the infrared one. 

\subsubsection{1034-293, $z=0.311$}
 
 It is polarized at $P_{opt}=13.8\pm1.18\%$ (Wills et al. 1992). The 
 limited infrared data give a variation of 1 mag in K and color indices 
 of (J-H) = 1.00$\pm$0.05, (J-K) = 1.95$\pm$0.15, and (H-K) = 0.95$\pm$0.20. 

\subsubsection{1055+018,$z=0.888$ }

 There are only two nights of infrared data for this polarized 
 ( $P_{opt}=6.\%$, Impey \& Tapia 1988) object  showing a variation of 
 $\Delta H = 0.54$ and color indices of (J-H) = 0.66$\pm$0.14, 
 (J-K) = 1.82$\pm$0.15, and (H-K) = 1.16$\pm$0.13.

\subsubsection{1156+295, $4C29.45, Ton599$, $z=0.728$}

 This variable ($\Delta m=5.0$, Branly et al. 1996) object shows high and 
 variable polarizations ($P_{IR}=28.06\% $, $P_{opt}$=28\%, Holmes et al. 
 1984; Mead et al. 1990). The optical polarization increase from $5\%$ to 
 20$\%$ between 1984 June 9 and 10 (Smith 1996). During the simultaneous 
 observations, 1156+295 did not show significant change in the shape of 
 the IR-UV spectrum when the flux varied (Glassgold et al. 1983).
 The infrared light curves show two double-peaked outbursts. There are 
 positive correlations between (J-K) and (J-H) as well as  (H-K), but the 
 association between color index and magnitude is complex, no definite 
 correlation can be obtained (see Fig. 8), which is similar to the result 
 found in the optical band by  Glassgold et al (1983)(also 
 See Fan 1999b).

\subsubsection{1218+304, $RS 4$, $z=0.130$}

 The X-ray source is one of the best observed in the ``2A'' catalogue of 
 high galactic-latitude sources  (Cooke et al. 1978; Wilson et al. 1979).  
 An optical polarization of $P_{opt}=6.6\%$ is known (Wills et al. 1992).
 There are 4 nights of infrared data showing a  variation of 
 $\Delta K = 0.91$ and color indices of  (J-H) = 0.67$\pm$0.10, 
 (J-K) = 1.50$\pm$0.29, and (H-K) = 0.78$\pm$0.16.

\subsubsection{1244-255, $z=0.638$}

 There are only 3 nights of  data showing a variation of 
 $\Delta H =  \Delta K = 0.95$ and color indices of  (J-H) = 0.86$\pm$0.25, 
 (J-K) = 1.69$\pm$0.22, and (H-K) = 0.82$\pm$0.27. An optical polarization 
 of  $P_{opt}=8\%-12\%$ (Impey \& Tapia 1988) and a variation of 
 $\Delta V=2.0$  (Bozyan et al. 1990) are reported.

\subsubsection{1253-055, $3C279$, $z=0.536$}

 3C279 is a well known member  of OVVs, a large optical variation of 
 $\Delta B\geq6.70$ mag (Eachus \& Liller 1975) and a highly  optical 
 polarization of $P_{opt}=44\%$ ( see Fan et al. 1996) and a possible
 variability period of 7-year (Fan 1999) are reported. It  
 is the prototype of superluminal radio sources and shows a violently optical 
  brightness increase of 2.0 mag during an interval of 24 hours 
 (Webb et al. 1990).  The infrared light curves show 3 outbursts. 
 The (J-K) is strongly associated with (J-H) and (H-K). There is a positive 
 correlation  between color index and magnitude ( see Fig. 9). 
 For (H-K) and J, there is a positive correlation between them when
 J is brighter than 14 magnitude but there is a negative correlation
 between them when the source is fainter than 14 magnitude. From
 above discussion, it is possible that the emission mechanism consists
 of two components in the case ( also see Fan 1999a,b).

\subsubsection{1510-089, $z=0.361$}

 1510-089 is one of the largest variable objects with  
 $\Delta B=5.4 (11.8-17.2)$ (Bozyan et al. 1990). A maximum  optical
  polarization of $P_{opt}=14.45\%$ is reported and the optical polarization 
 is found to be wavelength-dependent with $dP/\lambda > 0$(see Smith et al. 
 1988), the infrared polarization is $P_{IR} = 9.07\%$  (Mead et al. 1990). 
 The infrared light curves show a clear brightness increase with the  
 variation in the shorter wavelength being smaller than that in the longer 
 one. (J-K) is correlated with (H-K),  no correlated between  
 color index and magnitude can be found from the available data (see Fig.
 10).

\subsubsection{1641+395, $3C345$, $z=0.595$}
 
 3C345 is one of the well studied quasars (Bregman et al. 1986a; Smith 1996, 
 and references therein), it gives us a well-documented example of a clear 
 connection between brightness and polarization as well as polarization 
 angle (Smith 1996).  It is polarized at $P_{IR}=0\%-20.27\%$ 
 (Mead et al. 1990) and $P_{opt}=5\%-35\%$ (Bregman et al. 1986a).
 The optical variation of $\Delta m=2.5$ (Bregman et al, 1996a) is smaller 
 than that in the infrared variation of $\Delta J=3.16$ mag. The infrared 
 observations in the paper of Bregman et al (1986a) are also included 
 in our discussion although the paper is not listed in table 1 for they 
 have not indicated their telescopes used.  The infrared light curves show
  clearly two outbursts separating by about 11.3 years which is consistent
  with the 11.4-year optical outburst period ( Webb et al. 1988, also see 
 Fan et al. 1998a for summary). For the color indices, (J-K) is found 
 correlated strongly with (J-H) and (H-K). A correlation
 is found between color index and magnitude(see Fig. 11):
 (J-K) = 0.49 + 0.10H  with $r$ = 0.555 and $p = 7.6 \times 10^{-6}$,
 (J-K) = 0.27 + 1.76 (J-H)  with $r$ = 0.80 and $p = 6.6 \times 10^{-14}$,
 (J-K) = 0.003 + 1.93 (H-K)  with $r$ = 0.82 and $p = 1.0\times 10^{-14}$.

\subsubsection{1717+177, $OT129$, $z=?$}

 It is identified by Hoskins et al. (1974) and Condon et al. (1977) with 
  a starlike object of 18.5 mag (also see Veron \& Veron, 1993).
 There are only 4 nights of infrared data for this polarized 
 ($P_{opt.}=27\%$, Angel \& Stockman 1980; $P_{IR}=16.85\%$, Mead et al. 
 1990) object showing a variation of $\Delta K=1.89$ and color indices of 
 (J-H) =  0.92$\pm$0.03, (J-K) =  1.90$\pm$0.03, and (H-K) = 0.98$\pm$0.03.

\subsubsection{1722+119, $z=?$}

 This object is very interesting, its polarizations ($P_{opt}=17.6\pm1.0\%$ 
 and $P_{IR}=11.9\pm1.1\%$) are at a  typical level of a radio selected BL 
 Lac object (RBL) and are strongly wavelength-dependent 
 ($dP/d\lambda=-5.7\pm1.2\%$), the X-ray to optical luminosity ratio is 
 comparatively much higher and is similar to the values among XBLs
 (Brissenden et al. 1990). The 4 nights of  data show a variation of 1.0 mag 
 in J, H, and K and color indices of (J-H)=0.71$\pm$0.07, 
 (J-K)=1.38$\pm$0.06,  and (H-K)=0.67$\pm$0.04.

\subsubsection{1921-293, $OV-236$, $z=0.352$}
 
 An optical variation of  $\Delta m=2.64$ (Pica et al. 1988) and 
 polarizations of  $P_{opt.}=16.89\%$ and $P_{IR}=13.94\%$ 
 (Mead et al. 1990) are reported for this object. The  H light curve 
 shows a sharp brightness decrease,
  which leads H variation to be greater than the optical variation.  
 During the sharp decreasing period, there were no observations for J or K.  
 (J-K) is found correlated with  (J-H) and (H-K)
 (see Fig. 12):
  (J-K) = 0.63 + 1.30 (J-H)  with $r$ = 0.80 and $p = 9.1 \times 10^{-4}$,
 (J-K) = 0.17  + 1.80 (H-K)  with $r$ = 0.50 and $p = 4.7\% $.

\subsubsection{2155-304, $z=0.117$} 

 It is the brightest XBL in the UV.  The X-ray, UV,  and optical light 
 curves are well-correlated, suggesting a common origin of the optical 
 through X-ray emissions, and the X-rays lead the UV by $\sim$ 2-3 hours 
 (Edelson et al. 1995). It has been extensively studied in the UV and  
 X-rays (Pesce et al. 1996; Pian et al. 1996, and reference therein). 
 Polarizations of $P_{opt}=14.2\%$ ( Pesce et al. 1996) and 
 $P_{IR}=9.4\pm0.7\%$ (Mead et al. 1990) and an optical variation of 
 $\Delta m_{V} = 1.85$ ( Fan \& Lin 1999b) are reported. There is an
indication for 
 (J-K) to be correlated with (J-H) and (H-K) and H as well (see Fig. 13):
 (J-K) = 1.04 - 0.04H with $r$ = -0.37 and $p = 1.9\% $.

\subsubsection{2208-137,  $z=0.392$}
 
 The blazar is unusually showing symmetric double radio lobes 
 (Antonucci \& Ulvestad 1984). The double-lobed blazars are important in 
 establishing a kinship between blazars and normal double radio sources 
 (Antonucci et al. 1987).  The 7 nights of infrared data show a moderate 
 variation of 0.3 mag in the infrared and color indices of 
  (J-H) = 0.98$\pm$0.14, (J-K) = 2.10$\pm$0.14, and (H-K) = 1.11$\pm$0.01. 
 Polarizations of $P_{IR} = 9.32\%$ (Mead et al. 1990) and 
 $P_{opt} = 8.71\pm0.38\%$ (Moore \& Stockman 1981) are known.

\subsubsection{2223-052, $3C446$, ($z_{em}$=1.404)}

 The flat spectrum radio source is the prototype of the class of violently 
 variable quasars (Bregman et al. 1986b). During the optical outbursts, 
 the spectral index shows complex dependence on the brightness, sometimes 
 the slope of the optical continuous is steeper (Sandage et al. 1966 ) and 
 sometimes flatter (Visvanathan 1973) than during quiescent states.
 On occasions, week emission lines have been observed in the optical region, 
 the intensities of which do not change with the spectral flux 
 (Sandage et al 1966, see also Garilli \& Tagliferri 1986). Polarizations 
 of $P_{opt}=4\sim17.3\%$ (Impey \& Tapia 1990), $P_{IR}=16\%$ (Impey et al. 
 1982) and  $P_{10~GHz}=4\%$ (Mead et al. 1990) and an optical variation of 
 $\Delta m=5.0$  (Branly et al. 1996) are reported. The infrared light 
 curves shows a clear outburst, after which the brightness decrease 
 rapidly by about 4 mag in H. (J-K) is found correlated with (J-H) and (H-K),
  but the correlation between color index and magnitude is complex: There 
 is a positive connection between (H-K) and J but a negative correlation
 between (J-H) and K(see Fig. 14):
 (H-K) = 0.06 +0.06J with  $r$ = 0.589 and $p = 3.8 \times 10^{-4}$,
 (J-H) = 1.19 - 0.02K with  $r$ = -0.35 and $p = 3 \%$.

\subsubsection{2251+158, $3C454.3$, $z=0.859$}
 
 A polarization of $P_{opt} = 0\%-16.0\%$  and $\Delta m = 2.3$ are reported 
 in the paper of Angel \& Stockman (1980).  A variability of 0.5 mag over 
 a time scale of one day has also been observed in this object (Lloyd, 1984). 
 Smith et al (1988) observed this object on 1986 Dec. 23, 1987 June 23 and 
 24 and found the optical polarization is less than $6\%$ and the polarization
 is wavelength-dependent with $dP/d\lambda > 0$. The light curve indicates
 a variation of 1.57 mag in K which is smaller than in the optical one. (J-K) 
 is found to be correlated with (J-H) and (H-K) (see Fig. 15):
  (J-K) = 0.48 + 1.52 (J-H)  with $r$ = 0.94 and $p =   1 \times 10^{-4}$,
 (J-K) = -0.44 + 2.36 (H-K)  with $r$ = 0.86 and $p = 2.3 \times 10^{-3} $.

\subsubsection{2345-167, $z=0.576$}
 
  Smith et al.(1988) found that the optical polarization ($P_{B} = 4.78\%$, 
 $P_{R} = 11.81\%$, and $P_{I} = 11.90\%$)  wavelength-dependent. The maximum
 optical polarization  $P_{opt} = 3 \sim 19\%$ (Angel \& Stockman 1980) 
 and maximum optical variation of  $\Delta m = 2.55$ (Bozyan et al. 1990) 
 are reported. There  are only two nights of infrared observations 
 showing J = 16.54,  H = 15.61-15.85 and K = 14.58 and color indices of 
 (J-H) = 0.93$\pm$0.22, (J-K) = 1.96$\pm$ 0.25 and (H-K) = 1.03$\pm$0.21.

\subsection{Summary}

 In this paper, the infrared variations are presented for 30 blazars, the 
 light curves  are shown in Fig. 1 to 15  for 15 objects with enough 
 observations. The amplitude variation in the optical and infrared 
 bands are compared. 
 For the color-magnitude relation, some objects 
 (0215+015, 0422+004,  and 1641+395) 
 show a color  index increasing with magnitude, indicating that the 
 spectrum flattens  when the source brightens while 
 3C 279 shows a complex correlation between (H-K) and J meaning
 that the emission mechanism consists of two components, the rest
 objects do not show any  clear tendency between color  index 
 and magnitude. The color indexes suggest that the spectral
 indexes are in the range of $\alpha_{IR}$ = 0.77 to 2.37.

\begin{acknowledgements}{This work is supported by the National 
 Pan Deng Project of China and the National Natural Scientific
 Foundation of China. I thank Prof. R.G. Lin for his help 
 with the data and Dr. Y. Copin for his help with the figure} 
\end{acknowledgements}
\cite{}

\begin{small}
\begin{table}
\caption[]{Literature and telescopes for the data}
\begin{tabular}{cc}
\hline\noalign{\smallskip}
Observer(s) &  Telescope(s)\\ 
\hline\noalign{\smallskip}
Allen(1976)             &  UM/UCSD 1.5m \\            
Allen et al (1982)      &  A-A 3.9m; UKIRT 3.8m\\    
Bersonelli et al (1992) &  ESO 3.6m \& 2.2m\\   
Brindle et al (1986)    &  UKIRT 3.8m \\   
Brissenden et al(1990)  &  AUN 2.3m \\   
Brown et al (1989)      &  UKIRT 3.8m \\   
Cruz-Gonzales \& Huchra & CTIO 4m\\
 (1984)                 & KPNO 1.5m \\   
Cutri et al(1985)       & UOA 1.55m  \\   
Falomo et al (1993)     &  ESO 2.2m \\   
Falomo  (1990)          &  ESO 2.2m \\   
Garcia-Lario et al(1989) & TCS 1.5m  \\   
Gear      (1993)       &  UKIRT 3.8m  \\   
Gear et al (1985,1986)       &  UKIRT 3.8m  \\   
Glass (1981)            &  Sutherland 1.88m \\   
Glassgold et al(1983)   & UKIRT 3.8m \\
                        & Palomer Mt. 5m \\   
Holmes et al (1984)     &  UKIRT 3.8m  \\   
Impey et al (1982,1984)      &  UKIRT 3.8m \\   
Kidger \& Allan (1988)  & TCS 1.5m  \\   
Kidger \& Casares (1989) & TCS 1.5m  \\   
Kidger et al(1992)      & TCS 1.5m  \\   
Kidger et al (1993)     & TCS 1.5m            \\   
Kitilainen et al (1992) & UKIRT 4m \\  \hline

\end{tabular}
\end{table}

\begin{table}
\caption[]{Literature and telescopes for the data}
\begin{tabular}{cc}
\hline\noalign{\smallskip}
Observer(s) &  Telescope(s)\\ 
\hline\noalign{\smallskip}
Landau et al (1986)     &  UKIRT 3.8m; Hale 5m \&\\
                        &   Mount Lemnon 1.5m \\   
Ledden et al(1981)      & UM/UCSD 1.5m \\   
Lepine et al(1985)      &  ESO 3.6m \\   
Litchfield et al (1994) &  ESO 2.2m  \\   
Maraschi et al(1994)    &  Sutherland 1.9m  \\   
Massaro et al (1995)    & TIGRO 1.5m \& ESO 1.0m \\   
Mead et al (1990)       &  UKIRT 3.8m\\   
Neugebauer et al(1979)  & Hale 5.0m  \\   
O'Dell et al (1978)     &  UM/UCSD 1.5m \\   
Puschell \& Stein (1980)&  UM/UCSD 1.5m \\   
Rieke et al (1977,1979) &  UOA 90inch \& 61 inch \\   
Robson et al(1983)      &   UKIRT 3.8m         \\   
Robson et al(1988)      &   UKIRT 3.8m  \\   
Roelling et al (1986)   &  UKIRT 3.8m \\   
Sitko et al (1982,1983) & UM/UCSD 1.5m  \\   
Sitko \& Sitko (1991)   &  KPNO 1.3m \& 1.5m \\   
Smith et al (1987)      &  KPNO 2.1m \\   
Takalo et al(1992)      & TCS 1.5m\\   
Tanzi et al (1989)      &  ESO 1.5m \& 3.6m \\   
Worrall et al (1986)    &  MU/UCSD 1.5m \\   \hline

\end{tabular}
\end{table}

\begin{table*}
\caption[]{Observed Largest Variations of Blazars}
\begin{tabular}{cccccccccc}
\noalign{\smallskip}
\hline
\noalign{\smallskip}
 Name     & z & $A_{V}$ & $P_{opt}(\%)$ & $\Delta m_{opt}$ & $\Delta J$
& $\Delta H$& $\Delta K$  & (J-H) & (H-K)\\      
 \hline
0109+244  & -------   & 0.09  & 17.3  & 3.07 & 1.55       &  1.56   & 1.58  & 0.80 $\pm$ 0.22 &  0.87 $\pm$ 0.11 \\   
0215+015  & 1.715     & 0.0   & 20.   & 5.0  & 2.00       &  2.69   & 2.52  & 0.83 $\pm$ 0.05 &  0.81 $\pm$ 0.09 \\   
0323+022  &0.147      & 0.068 & 9.    & 1.3  & 0.30       &  0.52   & 0.97  & 0.73 $\pm$ 0.10 &  0.48 $\pm$ 0.10 \\    
0420-014  & 0.915     & 0.163 & 20.   & 2.8  & 2.46       &  2.88   & 2.61  & 0.80 $\pm$ 0.16 &  0.86 $\pm$ 0.16 \\               
0422+004  & ------    & 0.20  & 22.   & 2.2  & 1.69       &  3.25   & 3.41  & 0.82 $\pm$ 0.11 &  0.82 $\pm$ 0.07 \\               
0521-365  & 0.055     & 0.176 & 11.   & 1.4  & 0.74       &  0.89   & 1.25  & 0.80 $\pm$ 0.16 &  0.69 $\pm$ 0.21 \\               
0548-322  & 0.069     & 0.283 &       &      & 0.40       &  0.32   & 0.55  & 0.71 $\pm$ 0.09 &  0.45 $\pm$ 0.09 \\               
0716+332  & ------    & 0.405 & 7.4   &      & 2.06       &  2.05   & 2.01  & 0.73 $\pm$ 0.04 &  0.84 $\pm$ 0.12 \\               
0736+017  & 0.191     & 0.863 & 6.    & 1.35 & 2.28       &  2.07   & 2.71  & 0.81 $\pm$ 0.14 &  0.88 $\pm$ 0.13\\               
0823-223  & 0.91      & 0.60  & 11.   & 1.5  & 1.93       &  2.26   & 2.32  & 0.66 $\pm$ 0.13 &  0.65 $\pm$ 0.13 \\               
0912+297  & ------    & 0.05  & 13.   & 2.25 & 0.40       &  2.27   & 2.41  & 0.71 $\pm$ 0.08 &  0.85 $\pm$ 0.20 \\               
1156+295  & 0.728     & 0.0   & 28.   & 5.   & 4.47       &  3.82   & 3.97  & 0.84 $\pm$ 0.09 &  0.91 $\pm$ 0.09 \\     
1253-055  & 0.536     & 0.0   & 44.   & 6.7  & 4.57       &  4.26   & 4.45  & 0.93 $\pm$ 0.12 &  0.95 $\pm$ 0.09 \\   
1641+395  & 0.595     & 0.09  & 35.   & 2.5  & 3.16       &  3.13   & 3.15  & 0.87 $\pm$ 0.14 &  0.95 $\pm$ 0.10 \\               
1510-089  & 0.360     & 0.10  & 14.   & 5.4  & 0.88       &  0.97   & 1.25  & 1.14 $\pm$ 0.05 &  1.08 $\pm$ 0.09\\               
1921-293  & 0.352     & 0.47  & 17.   & 2.64 & 2.10       &  3.04   & 2.16  & 0.92 $\pm$ 0.08 &  0.90 $\pm$ 0.09 \\               
2155-304  & 0.117     & 0.0   & 14.2  & 1.85 & 1.26       &  1.88   & 1.24  & 0.66 $\pm$ 0.09 &  0.62 $\pm$ 0.07\\               
2223-052  & 1.404     & 0.01  & 17.3  & 5.0  & 3.84       &  3.96   & 3.77  & 0.93 $\pm$ 0.11 &  0.92 $\pm$ 0.17\\               
2251+158  & 0.859     & 0.10  & 19.   & 2.5  & 0.76       &  1.31   & 1.57  & 0.68 $\pm$ 0.11 &  0.84 $\pm$ 0.09\\ \hline
\end{tabular}
\end{table*}
\end{small}

\newpage
\begin{table*}
\caption[]{Near-Infrared Observations of 40 Radio-Selected BL Lac Objects}
\begin{tabular}{cccccccc}
\hline
\noalign{\smallskip}
 Name &  JD 2400000+ & J  & $\sigma$ J  & H  & $\sigma$ H & K  & $\sigma$
K \\
 (1)  &  (2)         & (3)& (4)         &(5) & (6)        &(7) & (8)  \\ 
\hline
\end{tabular}

\end{table*}

\newpage 
Figure Captions
	 
Fig.  1. Light curves and color index properties for 0109+224. $a$: J
light curve; $b$: H light curve; $c$: K light curve;
 $d$: (J-H) vs. K; $e$: J-K vs. K; $f$: (H-K) vs K; $g$: (H-K) vs. (J-H); $h$: J-K vs. (J-H); $i$: J-K vs. (H-K)\\
Fig.  2. Light curves and color index properties for 0215+015\\
Fig.  3. Light curves and color index properties for 0422+004\\
Fig.  4. Light curves and color index properties for 0521-365\\
Fig.  5. Light curves and color index properties for 0548-322\\
Fig.  6. Light curves and color index properties for 0736+017\\
Fig.  7. Light curves and color index properties for 0823-223\\
Fig.  8. Light curves and color index properties for 1156+295\\
Fig. 9. Light curves and color index properties for 1253-055\\
Fig. 10. Light curves and color index properties for 1510-089\\
Fig. 11. Light curves and color index properties for 1641+395\\
Fig. 12. Light curves and color index properties for 1921-293\\
Fig. 13. Light curves and color index properties for 2155-304\\
Fig. 14. Light curves and color index properties for 2223-052\\
Fig. 15. Light curves and color index properties for 2251+158\\

\newpage 

\begin{figure}
\epsfxsize=18cm
$$	 
\epsfbox{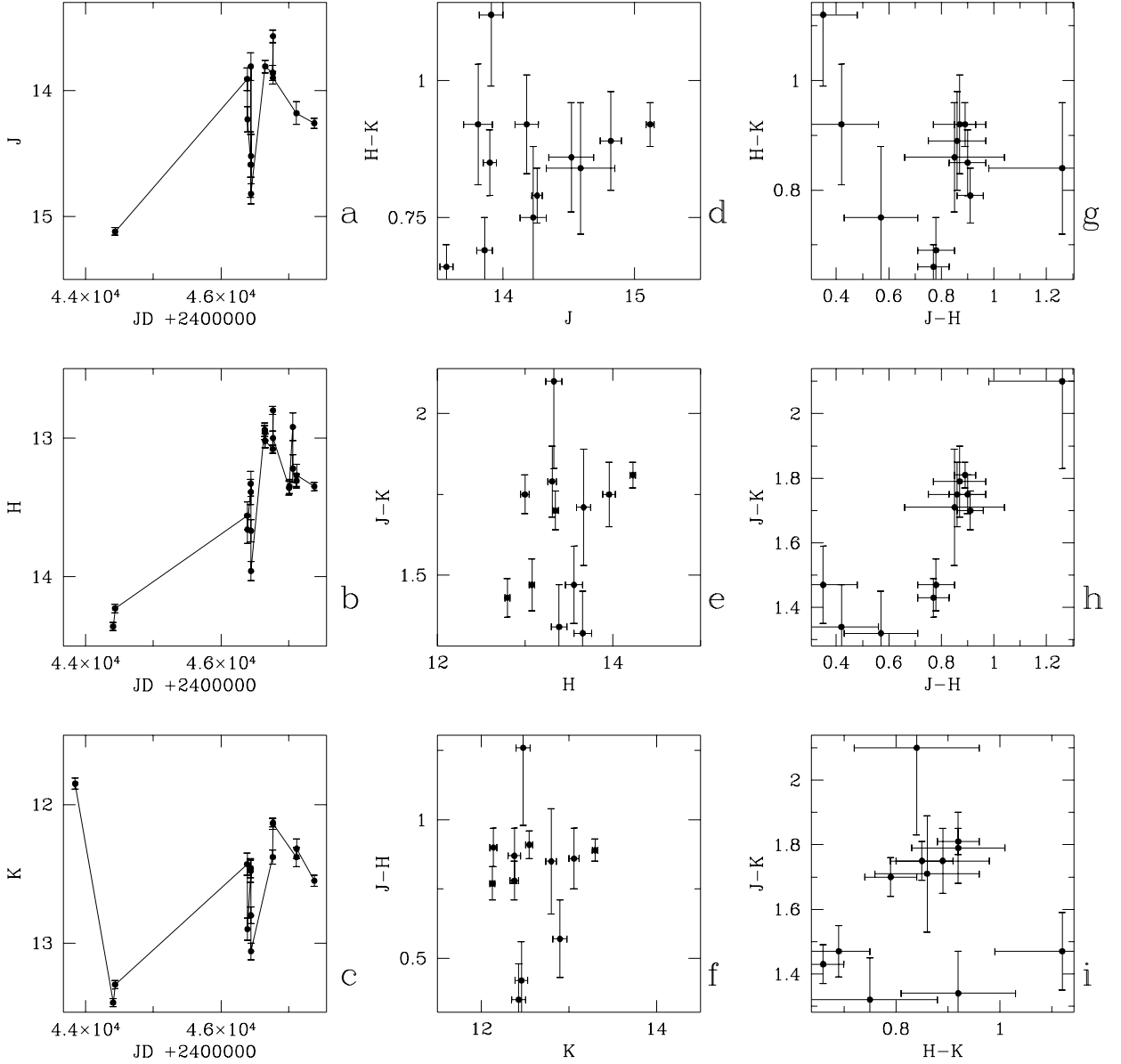}
$$	 
\caption {Light curves and color index properties for 0109+224. $a$: J light curve; $b$: H light curve; $c$: K light curve;
 $d$: (J-H) vs. K; $e$: (J-K) vs. K; $f$: (H-K) vs K; $g$: (H-K) vs. (J-H); $h$: (J-K) vs. (J-H); $i$: (J-K) vs. (H-K)}
\label{fig:1}
\end{figure}

\begin{figure}
\epsfxsize=18cm
$$	 
\epsfbox{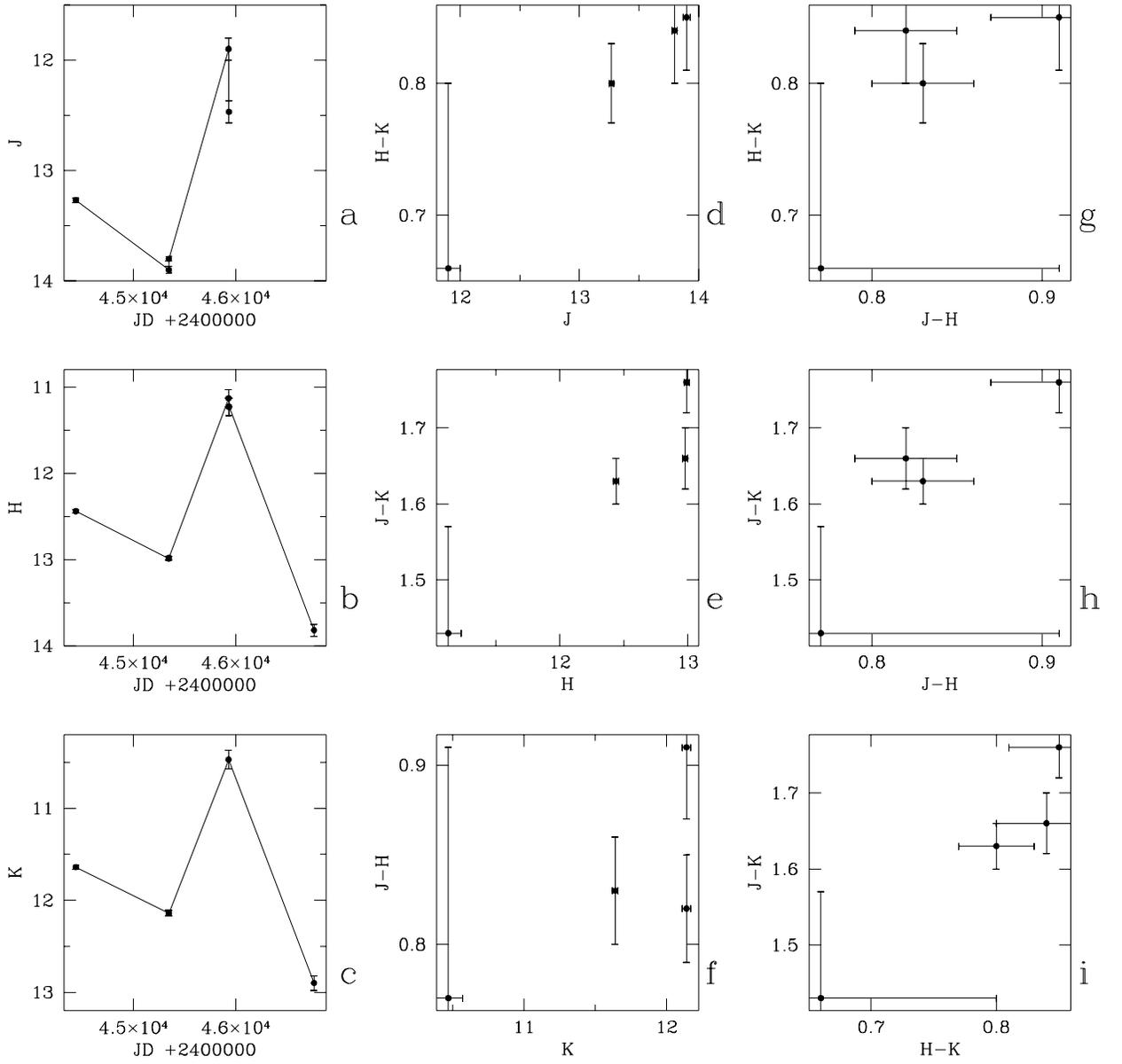}
$$	 
\caption {Light curves and color index properties for  0215+015.}
\label{fig:2}
\end{figure}
	 
\begin{figure}
\epsfxsize=18cm
$$	 
\epsfbox{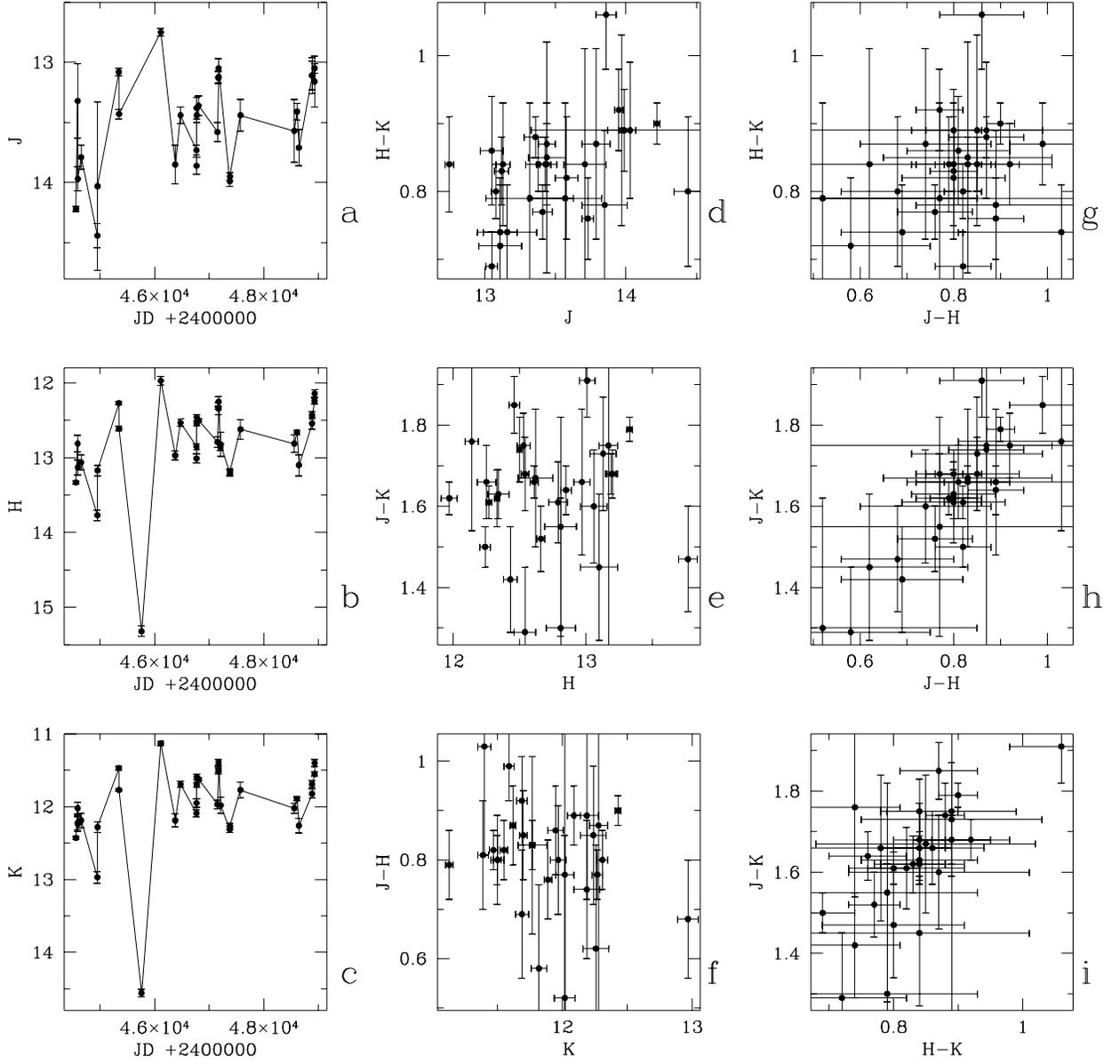}
$$	 
\caption {Light curves and color index properties for 0422+004 }
\label{fig:3}
\end{figure}

\begin{figure}
\epsfxsize=18cm
$$	 
\epsfbox{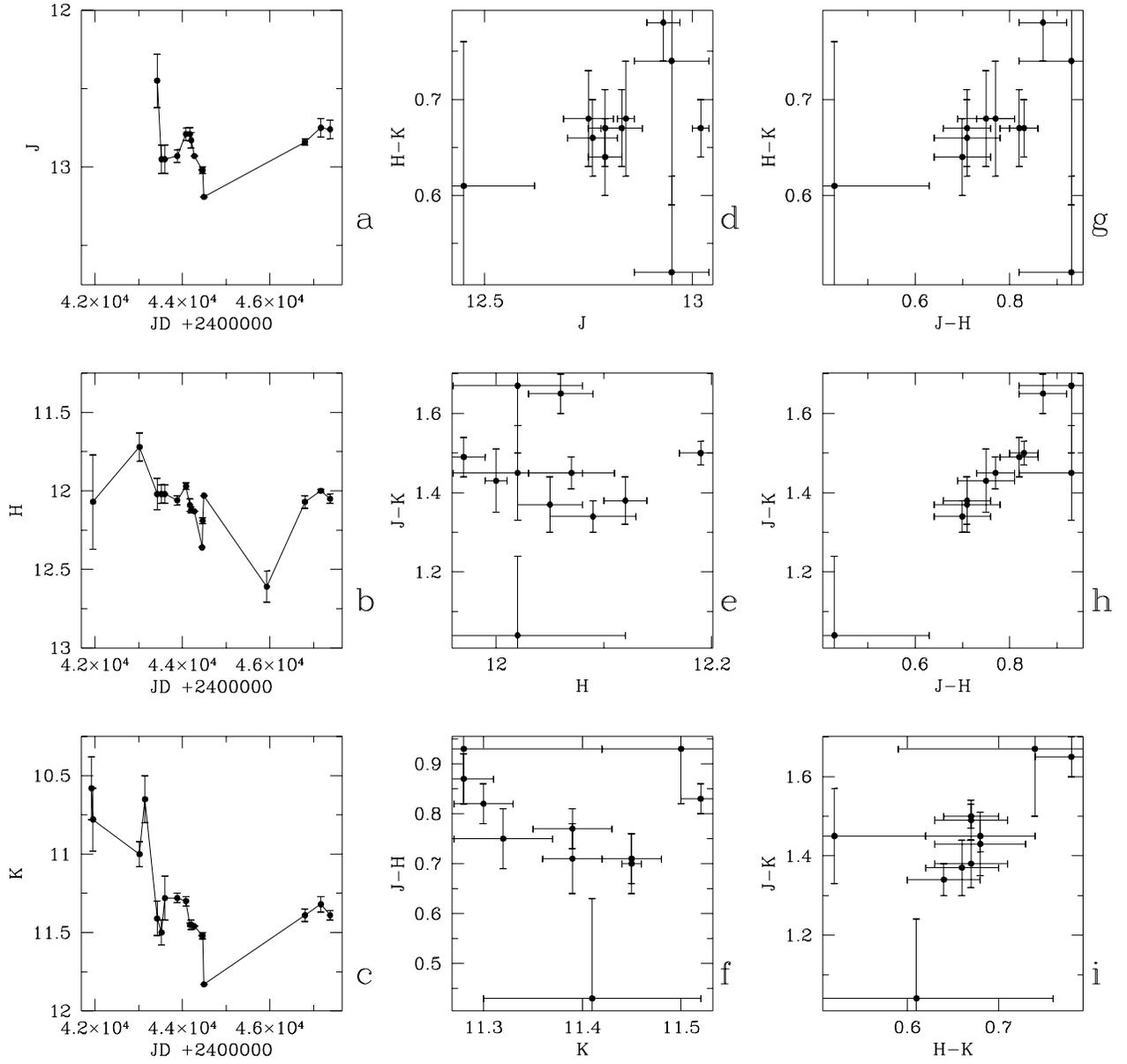}
$$	 
\caption {Light curves and color index properties for 0521-365 }
\label{fig:4}
\end{figure}

\begin{figure}
\epsfxsize=18cm
$$	 
\epsfbox{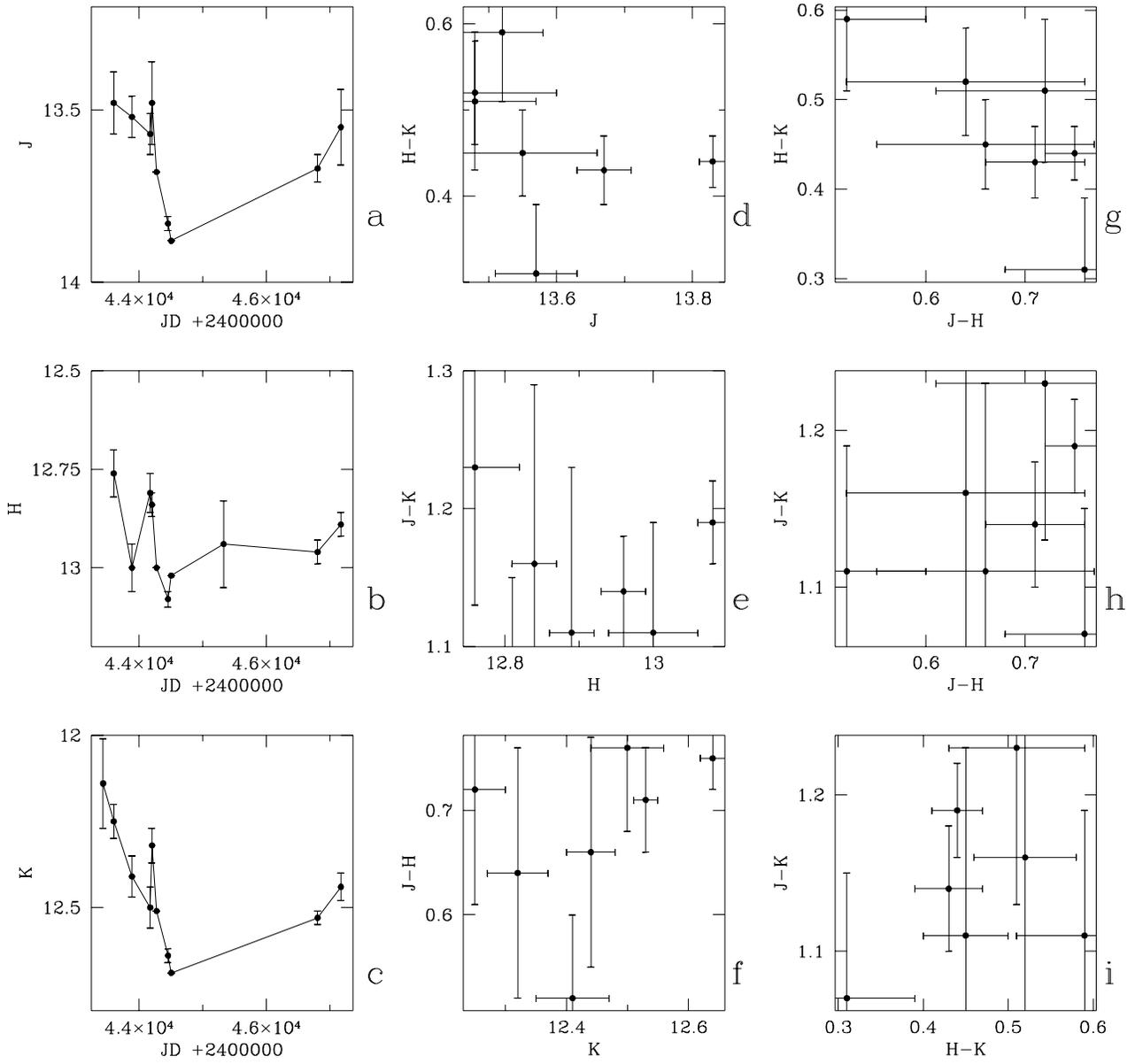}
$$	 
\caption {Light curves and color index properties for  0548-322}
\label{fig:5}
\end{figure}

\begin{figure}
\epsfxsize=18cm
$$	 
\epsfbox{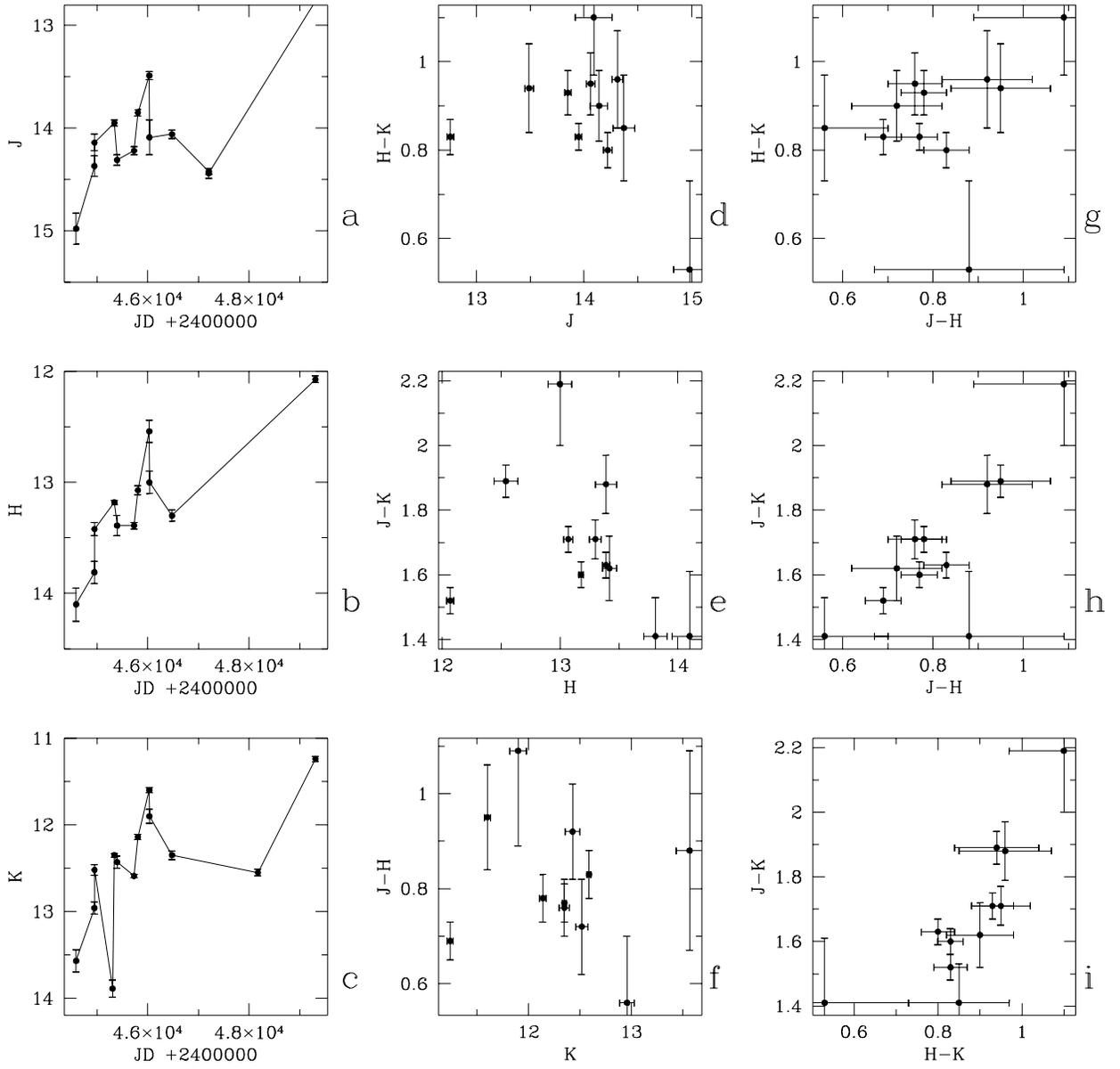}
$$	 
\caption {Light curves and color index properties for  0736+017}
\label{fig:6}
\end{figure}

\begin{figure}
\epsfxsize=18cm
$$	 
\epsfbox{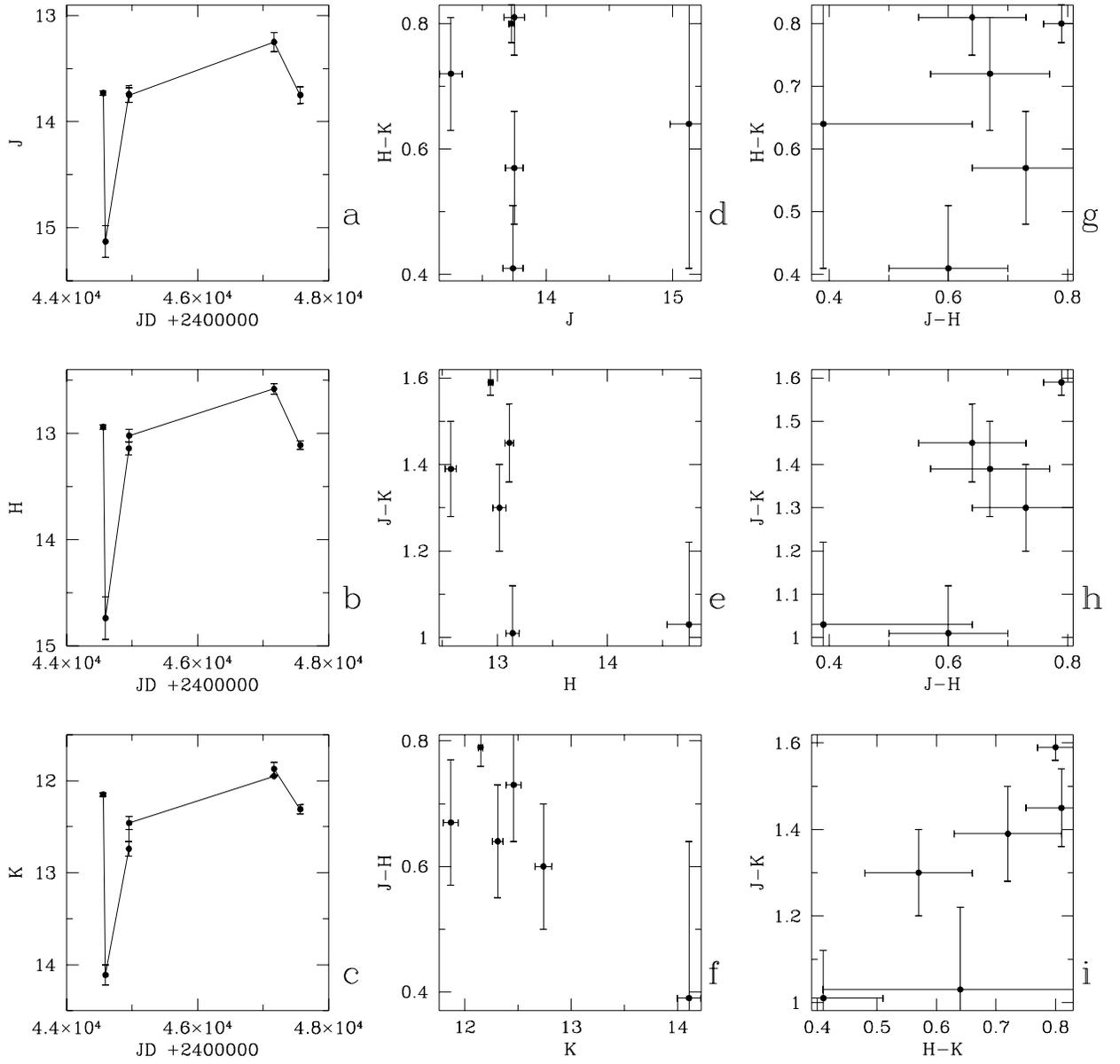}
$$	 
\caption {Light curves and color index properties for  0823-223}
\label{fig:7}
\end{figure}

\begin{figure}
\epsfxsize=18cm
$$	 
\epsfbox{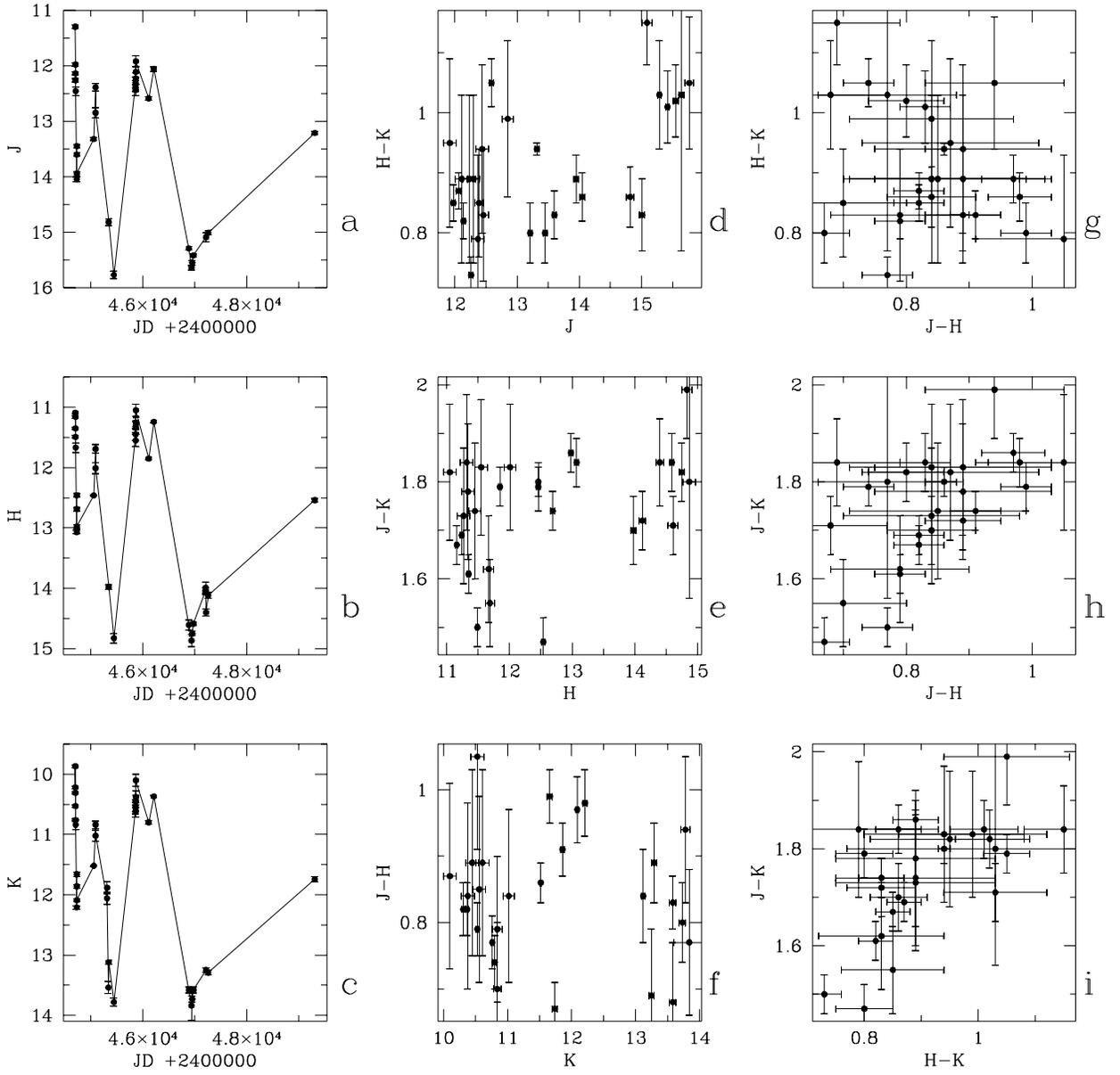}
$$	 
\caption {Light curves and color index properties for  1156+295}
\label{fig:8}
\end{figure}
	 
\begin{figure}
\epsfxsize=18cm
$$	 
\epsfbox{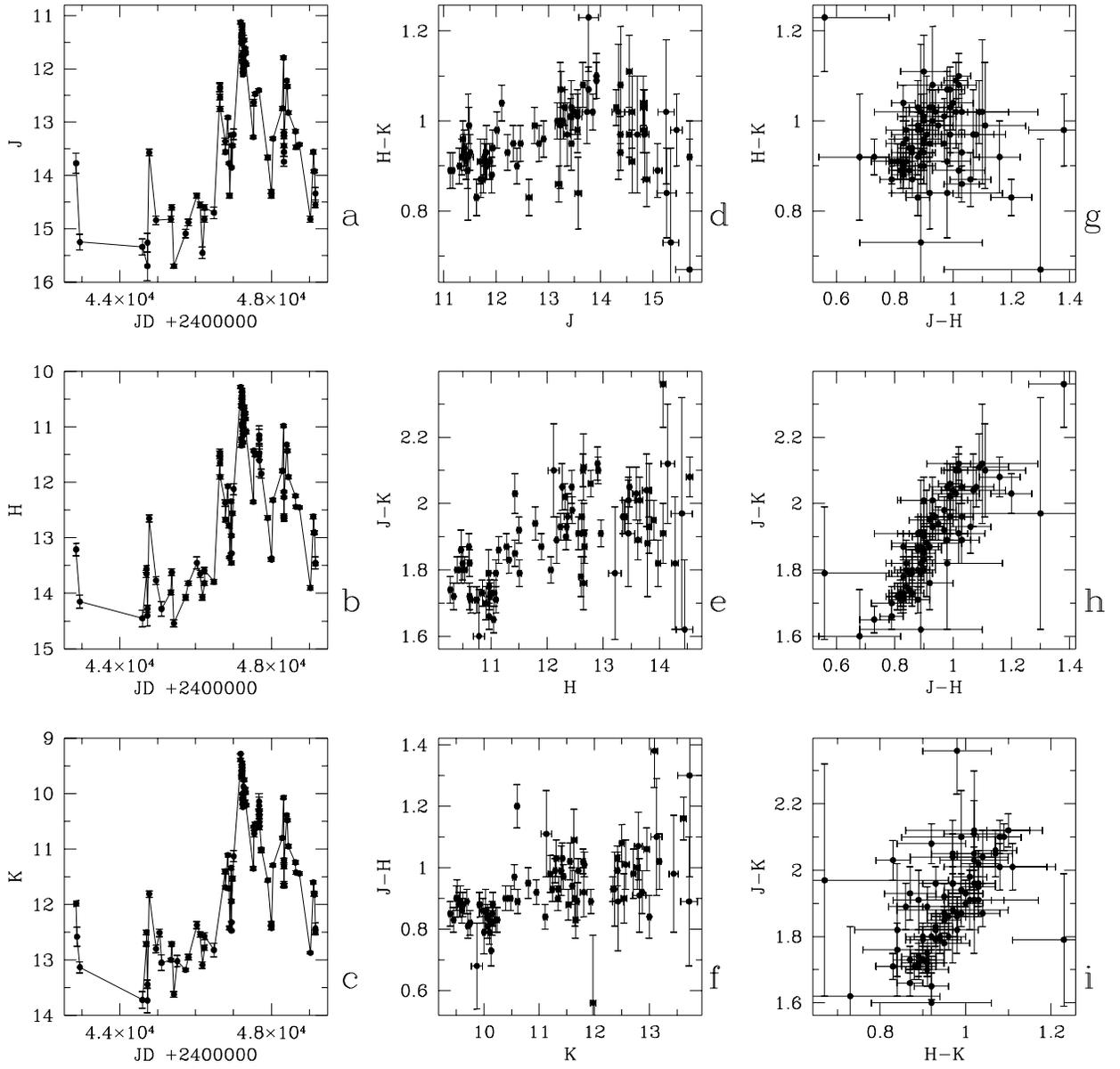}
$$	 
\caption {Light curves and color index properties for  1253-055}
\label{fig:9}
\end{figure}

\begin{figure}
\epsfxsize=18cm
$$	 
\epsfbox{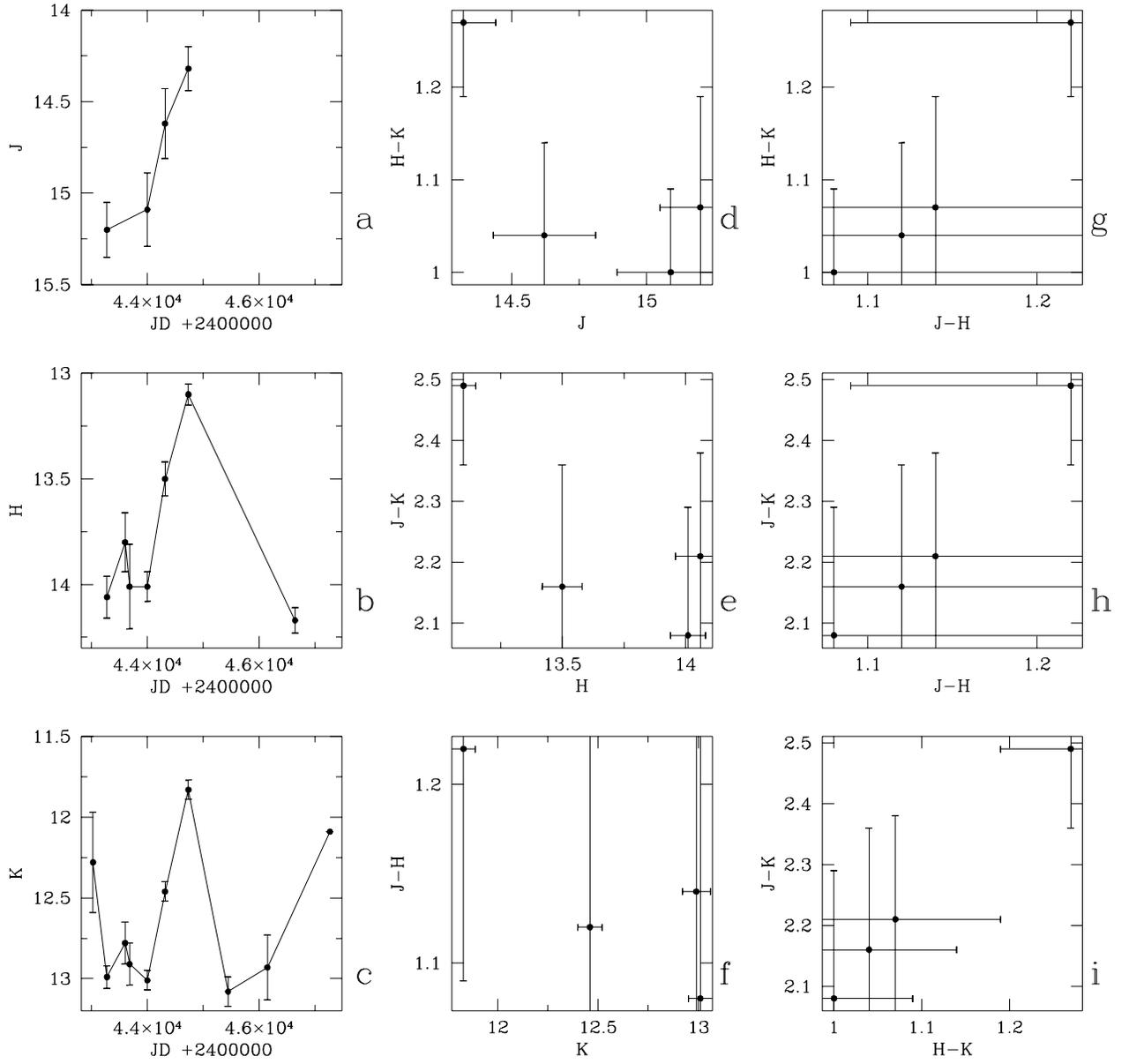}
$$	 
\caption {Light curves and color index properties for 1510-089}
\label{fig:10}
\end{figure}

\begin{figure}
\epsfxsize=18cm
$$	 
\epsfbox{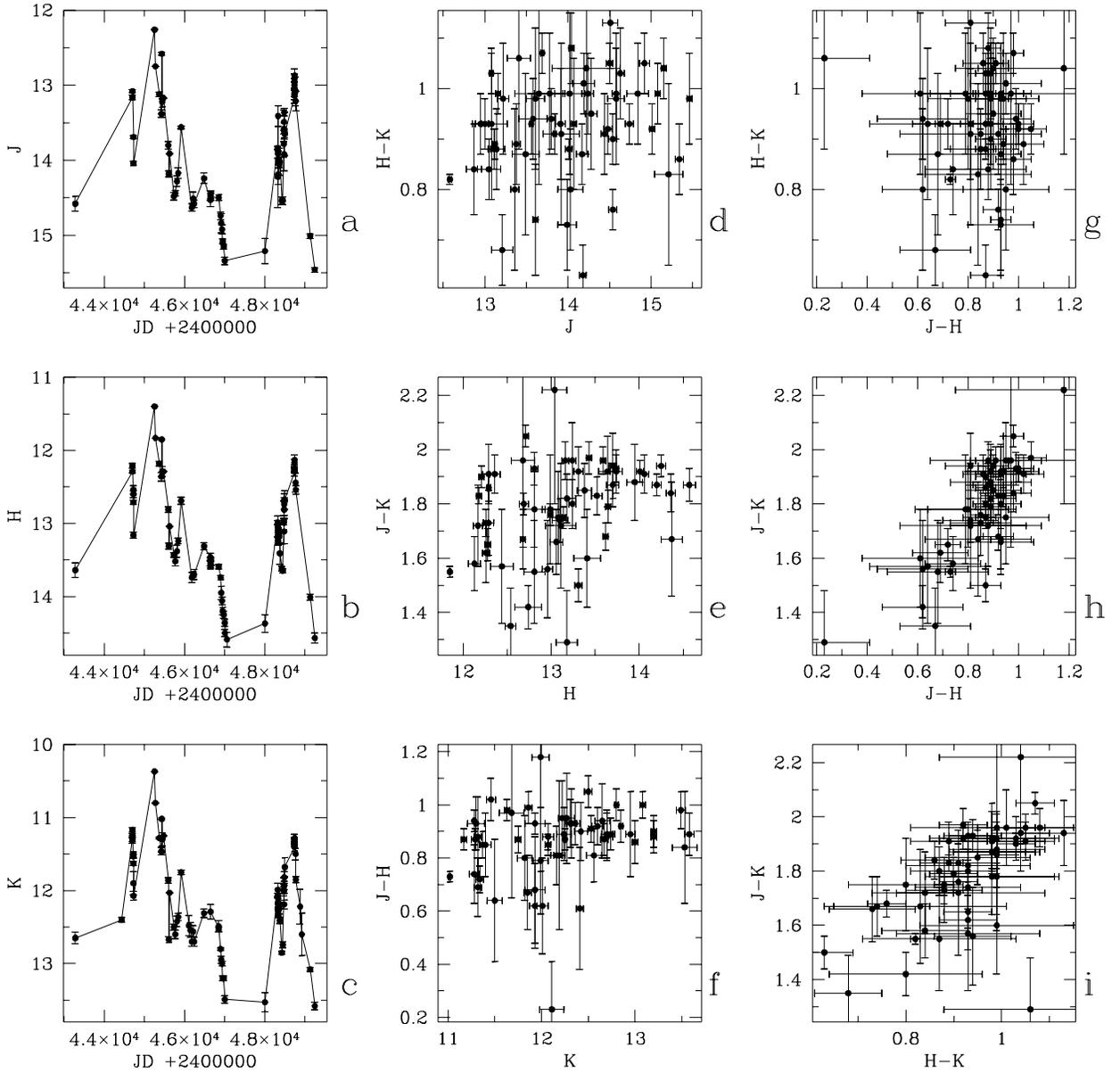}
$$	 
\caption {Light curves and color index properties for 1641+395}
\label{fig:11}
\end{figure}

\begin{figure}
\epsfxsize=18cm
$$	 
\epsfbox{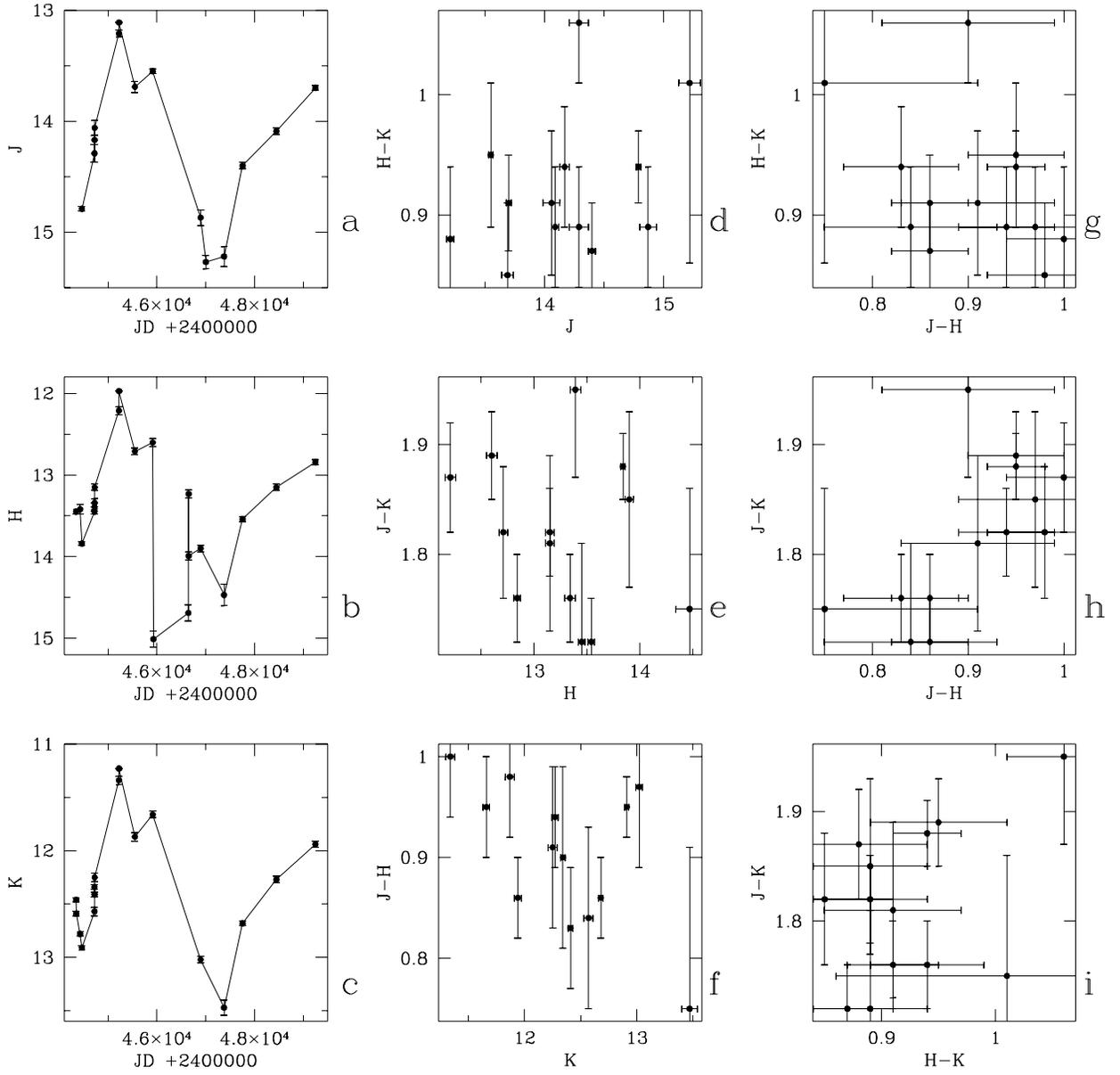}
$$	 
\caption {Light curves and color index properties for 1921-293}
\label{fig:12}
\end{figure}

\begin{figure}
\epsfxsize=18cm
$$	 
\epsfbox{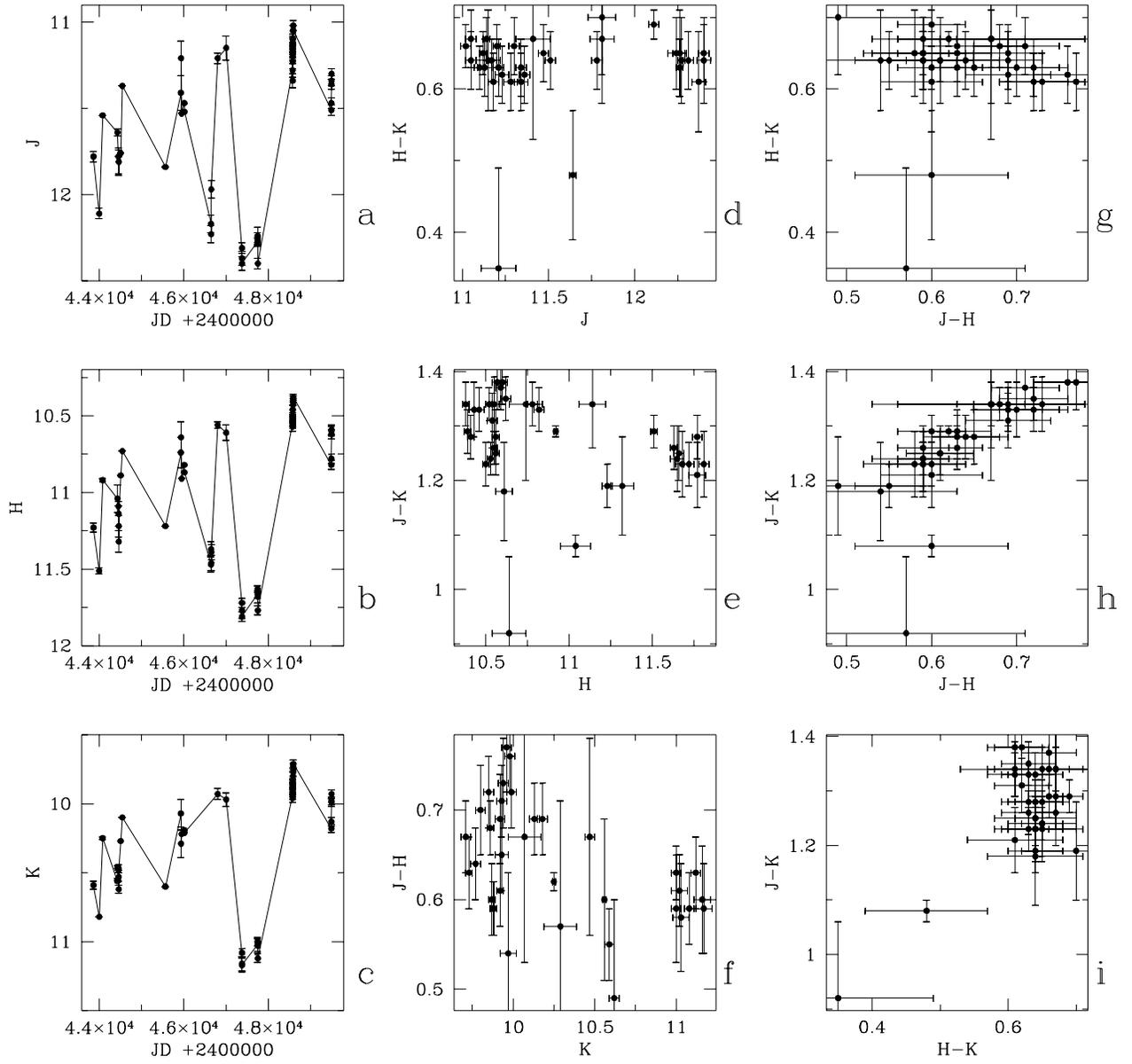}
$$	 
\caption {Light curves and color index properties for 2155-304}
\label{fig:13}
\end{figure}

\begin{figure}
\epsfxsize=18cm
$$	 
\epsfbox{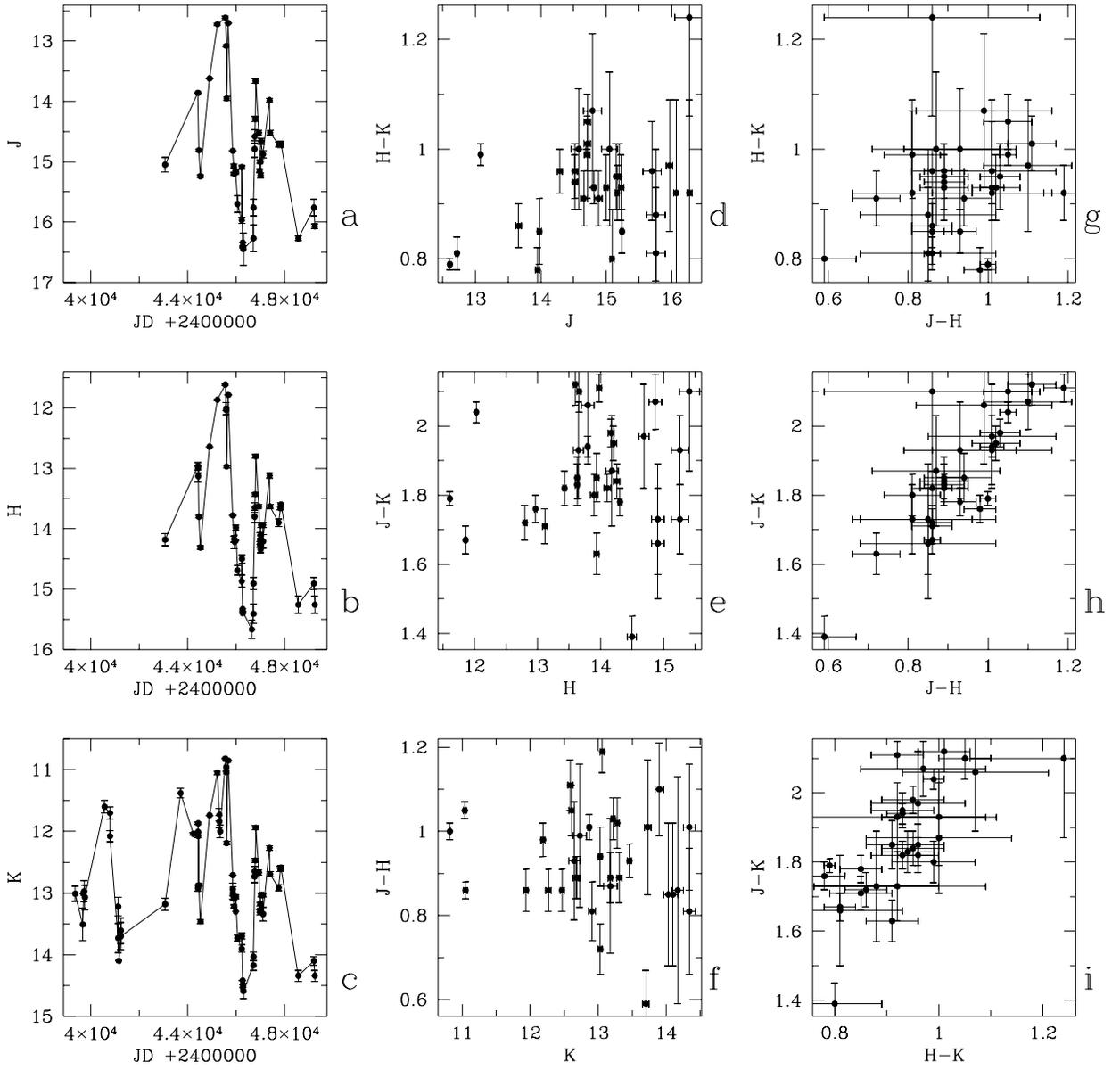}
$$	 
\caption {Light curves and color index properties for 2223-052}
\label{fig:14}
\end{figure}

\begin{figure}
\epsfxsize=18cm
$$	 
\epsfbox{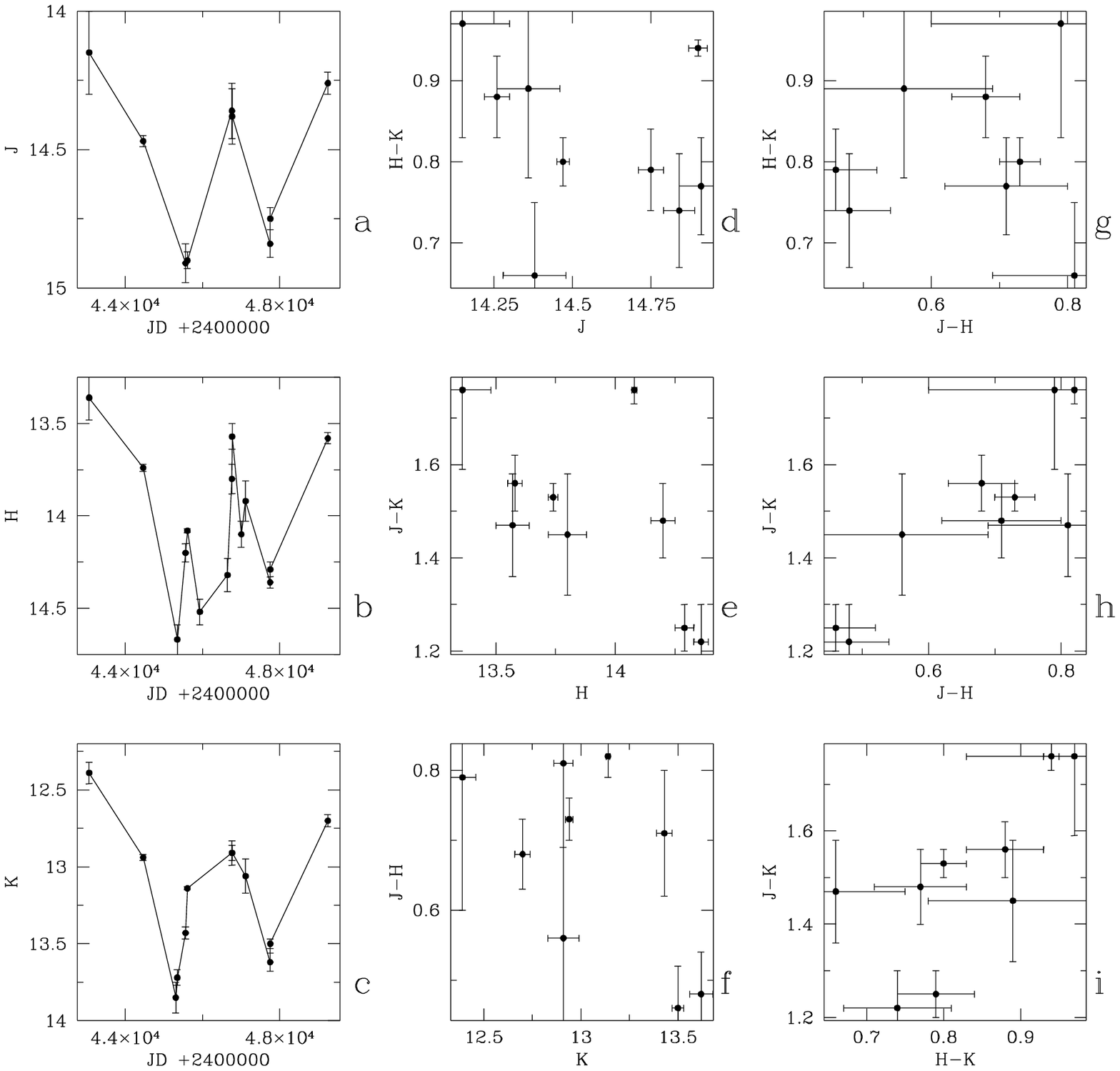}
$$	 
\caption {Light curves and color index properties for 2251+158}
\label{fig:15}
\end{figure}


\begin{thebibliography}{}
\bibitem[]{} Allen, D.A. 1976, ApJ, 207, 367
\bibitem[]{}  Allen D.A., Ward, M.J., Hyland, A.R.,  1982, MNRAS, 199, 969
\bibitem[]{} Angel J.R.P \& Stockman H.S. 1980, ARA\&A, 18, 321
\bibitem[]{} Antonucci R.R.J. et al. 1987, AJ, 93, 785
\bibitem[]{} Antonucci R.R.J. \& Ulvestad J.S. 1984, Nat. 308, 617.
\bibitem[]{} Bailey J. Hough J.H. \& Axon D.J. 1983, MNRAS, 203, 339.
\bibitem[]{} Bersannelli, M., Banchet, P, Falomo, R., \& Tanzi, E.G., 1992, AJ, 104, 28
\bibitem[]{}  Brindle C., Hough,J.H., Bailey, J.A., Axon, D.J., Hyland, A.R.,  1986, MNRAS, 221, 739
\bibitem[]{} Bolton J.G. Claarke M.E. \& Ekers, R.D.  1965b, Austrain J. Phys. 18, 627.
\bibitem[]{} Bolton J.G. Claarke M.E. Sandage A. \& Veropn P. 1965a, ApJ, 142, 1289
\bibitem[]{} Bozyan, E.P. et al. 1990, AJ, 99, 1421
\bibitem[]{} Branly R et al. 1996, ASP Conf. Ser. Vol. 110, p170
\bibitem[]{} Branly, R. Kilgard, R. Sadun, A. Shcherbanovsky, A. Webb, J. 1996, ASP Conf. Ser. Vol. 110, P170
\bibitem[]{} Bregman J.N. et al. 1986a, ApJ, 301, 708
\bibitem[]{} Bregman J.N. et al. 1986b, ApJ, 301, 698
\bibitem[]{} Brissenden R.J.V. Remillard R.A. Tuohy I.R. Schwartz D.A. Hertz, P.L. 1990, ApJ, 350, 578
\bibitem[]{} Brown, L.M.J. Robson, E.I., Gear, W.K., Smith, M.G.: 1989, ApJ, 340, 150
\bibitem[]{} Burbidge, E.M. et al. 1976, ApJ, 205, L117
\bibitem[]{} Condon J.J. Hicks P.D. \& Janncey D.L. 1977, AJ, 692
\bibitem[]{} Cooke B.A. et al. 1978, MNRAS, 182, 489
\bibitem[]{} Cruz-Gonzalez, I, \& Huchra, J.P. 1984, AJ, 89, 441
\bibitem[]{} Cutri R.M. et al. 1985, ApJ, 296, 423
\bibitem[]{} Danziger I.J., Fosbury, Goss, W.N. 1978,  Pitt. Conf. on BL Lacs. ed. A. Wolf. p204
\bibitem[]{} Disney M.J. 1974, ApJ, 193, L103
\bibitem[]{} Eachus L.A. \& Liller W. 1975, ApJ, 200, L61
\bibitem[]{} Falomo R. 1990, ApJ, 353, 114
\bibitem[]{} Falomo, R. et al. 1993, ApJ, 402, 532
\bibitem[]{} Fan J.H. 1999a, MNRAS (in press)
\bibitem[]{} Fan J.H. 1999b, A\&A, 347, 419 (astro-ph/9908080)
\bibitem[]{} Fan J.H. \& Lin R.G. 1999a, ApJS, 121, 131
\bibitem[]{} Fan J.H. \& Lin R.G. 1999b, A\&A, (accepted)
\bibitem[]{} Fan J.H., Xie G.Z., Pecontal E., et al. 1998a, ApJ 507,
173
\bibitem[]{} Fan J.H., Xie G.Z., Lin R.G., Qin Y.P.  1998b, A\&AS, 133,
216
\bibitem[]{} Fan J.H. 1997, Ap.L. \& Com. 35, 361
\bibitem[]{} Fugmann W. 1989, A\&A, 205, 86
\bibitem[]{} Garcia-Lario P. Kidger M.R. \& de Diego J.  et al. 1989 IAU Cir 4789
\bibitem[]{} Garilli B. \& Tagliferri G. 1986, ApJ, 301, 703
\bibitem[]{} Gear W.K. 1993, MNRAS, 264, 919.
\bibitem[]{} Gear W.K., Brown, L.M.J., Robson, E.I.,et al. 1986, ApJ, 304, 295
\bibitem[]{} Gear W.K. et al. 1985, ApJ, 291, 511
\bibitem[]{} Georganopoulos M \& Marsher A.P. 1996, ASP conf. Ser. Vol 110, P262
\bibitem[]{} Glass, I.S. 1981, MNRAS, 194, 795
\bibitem[]{} Glassgold et al. 1983 ApJ, 274, 101
\bibitem[]{} Holmes P.A., Brand, P.W.J.L., Wolstencroft, R.D., Williams, P.M.,  1984, MNRAS, 210, 961
\bibitem[]{} Hoskins D.G. et al. 1974 MNRAS, 166, 235.
\bibitem[]{} Impey C.D. Lawrence C.R. \& Tapia S. 1991, ApJ, 375, 461.
\bibitem[]{} Impey C.D. \& Neugebauer, G., 1988 , AJ, 95, 307
\bibitem[]{} Impey C.D. \& Tapia  1988, ApJ, 333, 666
\bibitem[]{} Impey C.D., Brand, P.W.J.L. Wolstencroft, R.D., Williams, P.M.,  1984 MNRAS, 209, 245
\bibitem[]{} Impey C.D., Brand, P.W.J.L. Wolstencroft, R.D., Williams, P.M.,  1982 MNRAS, 200, 19
\bibitem[]{} Kidger M.R. et al. 1993, ApJ, 407, L1
\bibitem[]{} Kidger, M.R, Takalo, L. Sillanpaa, A. 1992, A\&A, 264, 32
\bibitem[]{} Kidger M.R. \& Casares J. 1989 IAU Cir. 4739
\bibitem[]{} Kidger M.R. \& Allan P.M. 1988 IAU Circ. 4595
\bibitem[]{} Kitilainen, J.K. et al. 1992, MNRAS, 256, 125
\bibitem[]{} Kollgaard, R.J. 1994, Vista Astron. 38, 29
\bibitem[]{} Kuhr H. \& Schmidt G.D. 1990, AJ, 99, 1
\bibitem[]{}   Landau R.Golish, B., Jones, T.J., et al. 1986, ApJ, 308, 78
\bibitem[]{} Leacock R.L. Smith A.G. Edwaeds P.L. et al. 1976, ApJ, 206, L87
\bibitem[]{} Ledden J.E. O'Dell S.L. Stein W.A. Wisniewski W.Z. 1981, ApJ, 243, 47
\bibitem[]{} Lepine J.R.D. et al. 1985, A\&A, 149, 351
\bibitem[]{} Lin Y.C. Bertsch D.L. Dingus B.L. et al. 1997 ApJ, 476, L
\bibitem[]{} Lioyd, C. 1984, MNRAS, 209, 697
\bibitem[]{} Litchfield S.L. Robson E.I. Stevens J.A. 1994, MNRAS 270, 341
\bibitem[]{} Massaro E. et al. 1995, A\&A 299, 339
\bibitem[]{} Mead, A.R.G., Ballard, K.R., Brand, P.W.J.L. Hough, J.H., Brindle, C., Bailey, J.A., 1990, A\&AS, 83, 183
\bibitem[]{} Moore R.L. \& Stockman H.S. 1981, ApJ, 243, 60
\bibitem[]{} Neugebauer G et al. 1979, ApJ, 230, 79
\bibitem[]{} O'Dell, S.L. Pushell, J.J., Stein, W.A., Warner, J.W;, 1978, ApJS, 38, 267
\bibitem[]{} Pesce J.E. Urry C.M. Pian E. et al. 1996, ASP Conf. Ser. Vol. 110, P423
\bibitem[]{} Pettini M, Hunstead R.W., Murdoch, H.S. Blades J.C. 1983, ApJ, 273, 436
\bibitem[]{} Pian E. Urry C.M. Pesce J. et al. 1996, ASP Conf. Ser. Vol. 110, p417.
\bibitem[]{} Pica A.J. et al. 1988, AJ, 96, 1215
\bibitem[]{} Puschell, J.J. \& Stein, W.A. 1980, ApJ, 237, 331
\bibitem[]{} Rieke G.H. et al. 1979, ApJ, 232, L151
\bibitem[]{} Rieke G.H., Lebofsky, M.L., Kemp, J.C., Coyne, G.V., Tapia, S., 1977, ApJ, 218, L37
\bibitem[]{} Robson E.I. et al. 1983, Nat 305, 194
\bibitem[]{} Robson E.I. Smith M.G. Aycock J. \& Walther D.M. 1988 IAU Cir. 4556
\bibitem[]{} Roelling T.L., Beckin, E.E., Impey, C.D., Werner, M.W.,  1986, ApJ, 304, 646
\bibitem[]{} Sambruna R.M. Maraschi L. Urry C.M. 1996 ApJ, 463, 444
\bibitem[]{} Scarpa R. \& Falomo R. 1997, A\&A,  325, 109
\bibitem[]{} Sitko, M.L. Stein, W.A., Zhang, Y.X., Wisniewski, W.Z., 1983, PASP, 95, 724
\bibitem[]{}  Sitko M.L. \& Sitko A.K. 1991, PASP, 103, 160
\bibitem[]{} Sitko M.L. Stein, W.A. Zhang Y.X. \& Wisniewski W.Z. 1982  ApJ, 259, 486
\bibitem[]{} Smith, P.S. 1996, ASP Conf. Ser. Vol. 110, p135.
\bibitem[]{} Smith, P.S. Balonek, T., Elston, R. Heckert, P.A; 1987, ApJS, 64, 459
\bibitem[]{} Takalo L.O. et al. 1992, AJ, 104, 40
\bibitem[]{} Tanzi E.G. et al. 1989, BL Lac Objects, N334, p171
\bibitem[]{} Urry C.M. \& Padovani P. 1995, PASP, 107, 803
\bibitem[]{} Valtaoja E. et al. 1992, A\&A, 254, 80
\bibitem[]{} Veron-Cetty M.P.  \& Veron P. 1996, ESO Scientific Report No. 17
\bibitem[]{} Veron-Cetty M.P.  \& Veron P. 1993, A\&AS, 100, 521
\bibitem[]{} von Montigny C. et al. 1995, ApJ, 440, 525
\bibitem[]{} Webb J.R. Carini M.T., Clements, S. et al. 1990, AJ, 100, 1452
\bibitem[]{} Webb, J.R., Smith, A.G., Leacock, R.J. Fitzgibbons, G.L., Gombola, P.P., \& Sheppard, D.W.,  1988, AJ, 95, 374
\bibitem[]{} Wills, B.J. et al. 1992, ApJ, 398, 454
\bibitem[]{} Wilson A.S. Ward M.J. Axon D.J. Elvis M. Meurs E.J.A. 1979, MNRAS, 187, 109
\bibitem[]{} Worrall, D.M. et al. 1986, ApJ, 303, 589
\bibitem[]{} Zekl, H. et al. 1981, A\&A, 103, 342
\end{thebibliography}
\end{document}